\documentclass[a4paper,11pt]{article}

\usepackage{amsmath}
\usepackage{amsfonts}
\usepackage{amssymb}
\usepackage{graphicx}
\usepackage{subfig}    

\newcommand{\p}{\ensuremath{\partial}}

\newcommand{\Gam}{\ensuremath{\Gamma}}
\newcommand{\del}{\ensuremath{\delta}}

\newcommand{\eps}{\ensuremath{\varepsilon}}
\newcommand{\ep}{\ensuremath{\epsilon}}

\newcommand{\lam}{\ensuremath{\lambda}}
\newcommand{\Lam}{\ensuremath{\Lambda}}
\newcommand{\sig}{\ensuremath{\sigma}}

\newcommand{\Om}{\ensuremath{\Omega}}

\newcommand{\rmd}{\textrm{d}}

\newcommand{\Msol}{\ensuremath{M_{\rm sol}}}

\newcommand{\etal}{\emph{et al}.}

\renewcommand{\k}{\ensuremath{\mathbf{k}}}
\newcommand{\x}{\ensuremath{\mathbf{x}}}
\newcommand{\tW}{\ensuremath{\widetilde{W}}}
\newcommand{\R}{\ensuremath{\mathcal{R}}}

\newcommand{\Cite}[1]{Ref. \cite{#1}}
\newcommand{\Cites}[1]{Refs. \cite{#1}}
\newcommand{\eqn}[1]{Eqn. \eqref{#1}}
\newcommand{\eqns}[1]{Eqns. \eqref{#1}}
\newcommand{\fig}[1]{Fig. \ref{#1}}

\newcommand{\ph}[1]{\phantom{#1}}
\newcommand{\tab}[1]{Table \ref{#1}}

\newcommand{\be}{\begin{equation}}
\newcommand{\ee}{\end{equation}}
\newcommand{\beq}{\begin{IEEEeqnarray}}
\newcommand{\eeq}{\end{IEEEeqnarray}}
\newcommand{\la}[1]{\label{#1}}

\newcommand{\Cal}[1]{\ensuremath{\mathcal{#1}}}
\newcommand{\ti}[1]{{\tilde #1}}
\newcommand{\avg}[1]{\ensuremath{\langle \,#1\, \rangle}}
\newcommand{\delc}{\ensuremath{\delta_c}}
\newcommand{\Dt}{\ensuremath{\Delta t}}
\newcommand{\fnl}{\ensuremath{f_{\rm NL}}}
\newcommand{\gnl}{\ensuremath{g_{\rm NL}}}
\newcommand{\erfc}[1]{\ensuremath{{\rm erfc}\left(#1\right)}}
\newcommand{\erf}[1]{\ensuremath{{\rm erf}\left(#1\right)}}
\newcommand{\lamint}[1]{\ensuremath{\int_{-\infty}^{\infty}{\frac{d\lam}{\sqrt{2\pi}} #1}}} 

\begin{document}

\begin{center}
\Large{\textbf{An Improved Calculation of the
    Non-Gaussian Halo Mass Function}} \\[0.5cm]
 
\large{Guido D'Amico$^{\rm a,b}$, Marcello Musso$^{\rm c}$,
Jorge Nore\~na$^{\rm a,b}$, Aseem Paranjape$^{\rm c}$}
\\[0.5cm]

{\renewcommand{\thefootnote}{}\footnotetext[0]{E-mail: damico@sissa.it, musso@ictp.it, norena@sissa.it, aparanja@ictp.it}}

\small{
\textit{$^{\rm a}$ SISSA, via Bonomea 265, 34136 Trieste, Italy}}

\vspace{.2cm}

\small{
\textit{$^{\rm b}$ INFN - Sezione di Trieste, via Bonomea 265, 34136 Trieste, Italy}}

\vspace{.2cm}

\small{
\textit{$^{\rm c}$ Abdus Salam International Centre for Theoretical
  Physics\\ Strada Costiera 11, 34151, Trieste, Italy}} 

\end{center}

\vspace{.8cm}

\hrule \vspace{0.3cm}
\noindent \small{\textbf{Abstract}\\
The abundance of collapsed objects in the universe, or halo mass
function, is an important theoretical tool in studying the effects of
primordially generated non-Gaussianities on the large scale
structure. The non-Gaussian mass function has been calculated by
several authors in different ways, typically by exploiting the
smallness of certain parameters which naturally appear in the
calculation, to set up a perturbative expansion. We improve upon the
existing results for the mass function by combining path integral
methods and saddle point techniques (which have been separately
applied in previous approaches). Additionally, we carefully account
for the various scale dependent combinations of small parameters which 
appear. Some of these combinations in fact become of order unity for
large mass scales and at high redshifts, and must therefore be treated
non-perturbatively. Our approach allows us to do this, and also to 
account for multi-scale density correlations which appear in the
calculation. We thus derive an accurate expression for the mass
function which is based on approximations that are valid
over a larger range of mass scales and
redshifts than those of other authors. By tracking the terms ignored
in the analysis, we estimate theoretical errors for our result and
also for the results of others. We also discuss the complications
introduced by the choice of smoothing filter function, which we take
to be a top-hat in real space, and which leads to the dominant errors
in our expression. Finally, we present a detailed comparison between
the various expressions for the mass functions, exploring the accuracy
and range of validity of each.\\
\noindent
\hrule
\vspace{0.5cm}
}
\setcounter{footnote}{0}

\section{Introduction}

The primordial curvature inhomogeneities, generated by the
inflationary mechanism, obey nearly Gaussian statistics. 
The deviations from Gaussianity, while expected to be small,
provide a unique window into the physics of inflation.
For example, single-field slow-roll models of inflation lead to a
small level of non-Gaussianity (NG), so that an observation of a large 
NG would indicate a deviation from this paradigm.  

Until a few years ago, the main tool to constrain NG
was considered to be the statistics of the cosmic microwave background
(CMB) temperature field, since inhomogeneities at the CMB epoch are
small and their physics can be described by a perturbative treatment. 
In recent years, however, thanks to observations and developments in
the theory, the large-scale structure (LSS) of the universe has
emerged as a complementary probe to constrain primordial
NG. While it is true that the $n$-point functions of the 
density field on small scales are dominated by the recent
gravitational evolution, and do not reflect anymore the statistics of 
primordial perturbations, it turns out that the abundance of very
massive objects, which form out of high peaks of the density
perturbations, is a powerful probe of primordial NG. 
In this context, much attention has been given recently to three
possible methods of constraining the magnitude and shape of the
primordial NG with the LSS: the galaxy power spectrum,
the galaxy bispectrum and the mass function. It was pointed out in
\Cites{Dalal:2007cu, Matarrese:2008nc} that a NG of a local type induces a
scale dependence on the galaxy power spectrum, thus making it a
sensitive probe of the magnitude of local NG
$f_{NL}^{loc}$. From \Cite{Slosar:2008hx} one finds the following
constraints: $-29 < f^{loc}_{NL} < +69$, already comparable with those
obtained from CMB measurements in \Cite{Komatsu:2010fb}: $-10 <
f^{loc}_{NL} < +74$. The future is even more promising, with
precisions of $\Delta f^{loc}_{NL} \sim 10$~\cite{Sartoris:2010cr} and
$\Delta f^{loc}_{NL} \sim 1$~\cite{Carbone:2008iz,Carbone:2010sb,Cunha:2010zz} being claimed for
future surveys. The galaxy bispectrum is also a promising probe of
NG as it could be more sensitive to other triangle
configurations~\cite{Sefusatti:2009qh}. The mass
function -- which is the focus of this work, and which we discuss in
detail below -- has been used for example in~\Cite{Sartoris:2010cr}
together with the scale dependent bias to produce forecasts for future
surveys, and in~\Cite{Jimenez:2009us} in an attempt to explain the
presence of a very massive cluster at a large redshift as an
indication of a large NG. For more references and
information we refer the reader to reviews summarizing recent results 
on these topics~\cite{Verde:2010wp, Desjacques:2010jw}.

The formation of bound dark matter halos from initially small density
perturbations, as seen in numerical simulations, is a complicated and
violent process. Some insight into the physics involved has been
gained from the study of analytical models. The quantity of interest
is the halo mass function, defined as the number density of dark
matter halos with a mass between $M$ and $M + dM$,
\be
\frac{dn}{dM} = \frac{\bar\rho}{M^2}f(\sig)\left| \frac{d\ln\sig}{d\ln
M} \right|\,,
\la{mass-func-conventional}
\ee
where $\bar{\rho}$ is the average density of the universe, 
$\sigma(M)$ is the variance of the density contrast $\del_R$
filtered on some comoving scale $R$ corresponding to the mass $M$, and
the function $f(\sigma)$ is to be computed. Throughout this work, we
will refer to $f(\sigma)$ itself as the mass function. A very useful
tool in the analysis is the spherical collapse
model~\cite{Gunn:1972sv}, which predicts that the value of the
linearly extrapolated density contrast of a spherical halo, at the
time when the halo collapses, is $\delc\simeq1.686$, with a weak
cosmology dependence. This value serves as a collapse threshold for
determining which inhomogeneous regions will end up as collapsed
objects. Using this idea, Press \& Schechter~\cite{Press:1973iz}
(PS) first computed the mass function $f(\sigma)$ in the case of
Gaussian initial conditions. Their calculation
however suffered from a problem of undercounting which affects the
overall normalization -- their approach does not count underdense
regions embedded in larger overdense regions as eventually collapsed
objects. To account for this discrepancy, PS introduced an ad-hoc
factor of $2$ by demanding that the mass function be correctly
normalized, such that all the mass in the universe must be contained
in collapsed objects. In the excursion set approach, Bond
\etal~\cite{Bond:1990iw} resolved this issue and derived a correctly
normalized mass function, for Gaussian initial conditions. They argued 
that the filtered density contrast $\del_R$ follows a random walk as a
function of the filtering scale, and the problem of computing
$f(\sigma)$ is translated into the problem of finding the rate of
``first crossing'' of the barrier $\delc$, whose solution is
well-known. We will study this formalism in detail in
section~\ref{calc-f} for the more general non-Gaussian case.  

Turning to non-Gaussianities, the most popular non-Gaussian mass 
functions are those due to Matarrese, Verde and
Jimenez~\cite{Matarrese:2000iz} (MVJ) and
LoVerde~\etal~\cite{LoVerde:2007ri} (LMSV). Both groups used 
the PS approach, by modifying the probability density function for the 
(linearized) density contrast to describe non-Gaussian initial
conditions. In their prescription, the relevant object is the
\emph{ratio} $R_{\rm ng}$ of non-Gaussian to Gaussian mass
functions. The full mass function is usually taken as the product of
$R_{\rm ng}$ and an appropriate Gaussian mass function as given by
$N$-body simulations, e.g. the Sheth \& Tormen mass
function~\cite{Sheth:2001dp}. It is not clear however that this is the
correct way to proceed. Indeed, in a series of
papers~\cite{Maggiore:2009rv,Maggiore:2009rw,Maggiore:2009rx}, 
Maggiore \& Riotto (MR) presented a rigorous approach to the
first-passage problem in terms of path integrals, and
in~\Cite{Maggiore:2009rx} they pointed out that a PS-like prescription 
in fact misses some important non-Gaussian effects stemming from
$3$-point correlations between \emph{different} scales (so-called
``unequal time'' correlators).  

On the other hand, MR treated non-Gaussian contributions to $f(\sig)$
by simply linearizing in the $3$-point function of $\del_R$, i.e. by
linearizing in the non-Gaussian parameter \fnl. Since the
NG are assumed to be small, in the sense that the
parameter $\ep = \langle\delta^3\rangle/\sigma^{3}$ satisfies
$\ep\ll1$, one might expect that such a perturbative treatment is
valid. However, another  crucial ingredient in the problem is that the
length scales of interest are large, which leads to a \emph{second}
small parameter $\nu^{-1}$ where $\nu = \delc/\sig$. This is evident
in the calculations of MR, who crucially use $\nu^{-2}\propto\sig^2$
as a small parameter. Any perturbative treatment now depends not only
on the smallness of $\ep$ and $\nu^{-1}$ individually, but also on the
specific combinations of these parameters which appear in the
calculation. It is known (and we will explicitly see below) that a
natural combination that appears is $\ep\nu^3$, which can become of
order unity on scales of interest. The mass functions given by LMSV and
MR therefore break down as valid series expansions when this
occurs. Interestingly, MVJ's PS-like treatment on the other hand
involved a saddle point approximation,  allowing them to
\emph{non-perturbatively} account for the $\ep\nu^3$ term (which
appears in an exponential in their approach). For a discussion, see
\Cite{Valageas:2009vn}.

It appears to us therefore, that there is considerable room for
improvement in the theoretical calculation of the mass function. The
goal of our paper is twofold. Firstly, we present a rigorous
calculation of the mass function in the following way : (a) we use the
techniques developed by MR
in~\Cites{Maggiore:2009rv,Maggiore:2009rw,Maggiore:2009rx}, which
allow us to track the complex multi-scale correlations involved in the
calculation, and (b) we demonstrate that MR's approach can be combined
with saddle point techniques (used by MVJ), to non-perturbatively
handle terms which can become of order unity.  This leads to an
expression for the mass function which is valid on much larger scales
than those presented by MR and LMSV.  Secondly, by keeping track of
the terms ignored, we calculate theoretical error bars on the
expressions for $f(\sig)$ resulting not only from our own
calculations, but also for those of the other
authors~\cite{Matarrese:2000iz,LoVerde:2007ri,Maggiore:2009rx}. Since
the terms ignored depend on $\nu$ in general, these error bars are
clearly scale dependent. This allows us to estimate the validity of
each of the expressions for the mass function at different scales, but
importantly it also allows us to analytically compare between
different expressions.  In this paper we will not explicitly account
for effects of the ellipsoidal collapse model
\cite{Bardeen:1985tr,Sheth:1999su}, since these are expected to be
negligible on the very large scales which are of interest to us. For a
recent treatment of ellipsoidal collapse effects in the presence of
non-Gaussianities on scales where $\ep\nu^3\ll1$, see Lam \& Sheth
\cite{Lam:2009nb}. For a different approach to computing the
non-Gaussian halo mass function, see \Cite{Dalal:2007cu}, where the
authors proposed that this mass function can be approximated as a
convolution of the \emph{Gaussian} mass function with a probability
distribution function that maps between halos identified in Gaussian
and non-Gaussian $N$-body simulations. This probability distribution
itself was approximated as a Gaussian, with mean and variance fit
from simulations. While in practice this approach is easy to
implement, it relies heavily on the output of $N$-body
simulations. Our approach, on the other hand, allows us to compute a
mass function almost entirely from first principles.

This paper is organized as follows. In section~\ref{models} we
fix some notation and briefly introduce the two most popular shapes of
primordial NG, i.e. the local and equilateral ones.
In section~\ref{calc-f} we present our calculation
of the mass function.
In section~\ref{consistency} we discuss
certain subtleties regarding the truncation of the perturbative
series, and also compare with the other expressions for $f(\sig)$
mentioned above.
In section~\ref{diffuse-filter} we discuss the effects
induced by some additional complications introduced in the problem due
to the specific choice of the filter function~\cite{Maggiore:2009rv},
which we take to be a top-hat in real space, and due to the inclusion of
stochasticity in the value of the collapse threshold
$\delc$  (which is also expected to partially account
for effects of ellipsoidal collapse)~\cite{Maggiore:2009rw} . In
section~\ref{results} we compare our final
result~\eqn{f-fullcorrected} with those of other authors, 
including  theoretical errors for each, and conclude with a brief
discussion of the results and directions for future work. Some
technical asides have been relegated to the Appendices.

\section{Models of non-Gaussianity}
\la{models}
\noindent
We need to relate the linearly evolved density field to the
primordial curvature perturbation, which carries the information of
the non-linearities produced during and after inflation. We start from
the Bardeen potential $\Phi$ on subhorizon scales, given by
\be
\Phi(\k, z) = - \frac{3}{5} T(k) \frac{D(z)}{a} \R(k) \, ,
\la{Phi}
\ee
where $\R(\k)$ is the (comoving) curvature perturbation, which stays
constant on superhorizon scales; $T(k)$ is the transfer function of
perturbations, normalized to unity as $k \to 0$, which describes the
suppression of power for modes that entered the horizon before the
matter-radiation equality; and $D(z)$ is the linear growth factor of
density fluctuations, normalized such that $D(z)=(1+z)^{-1}$ in the
matter dominated era. Then, the density contrast field is related to
the potential by the Poisson equation, which in Fourier space reads
\begin{align}
\del(\k, z) &= - \frac{2 a k^2}{3 \Om_{m} H_0^2} \Phi(\k,z)
= \frac{2 k^2}{5 \Om_{m} H_0^2} T(k) D(z) \R(k) \nonumber \\
&\equiv \Cal{M}(k,z) \R(k) \, ,
\la{transfer}
\end{align}
where we substituted \eqn{Phi}.
Here, $\Om_{m}$ is the present time fractional density of matter (cold
dark matter and baryons), and $H_0=100h\,{\rm km\,s}^{-1}{\rm
  Mpc}^{-1}$ is the present time Hubble constant. The redshift
dependence is trivially accounted for by the linear growth factor
$D(z)$ and in the following, for notational simplicity, we will often
suppress it. All our calculations will use a reference \Lam CDM
cosmology compatible with WMAP7 data~\cite{Komatsu:2010fb}, using
parameters $h=0.702$, $\Om_{m}=0.272$, present baryon density $\Om_b=0.0455$,
scalar spectral index $n_s=0.961$ and $\sig_8=0.809$, where
$\sig_8^2$ is the variance of the density field smoothed on a
length scale of $8h^{-1}$Mpc. For simplicity, for the transfer function $T(k)$ we use
the BBKS form, proposed in Bardeen~\etal~\cite{Bardeen:1985tr}:
\be
T_{\rm BBKS}(x) \equiv \frac{1}{2.34x}\ln\left(1+2.34x\right) \left(
1+3.89x+(16.1x)^2 +(5.46x)^3 +(6.71x)^4\right)^{-1/4} \,,
\la{bbks}
\ee
where $x\equiv k(h{\rm Mpc}^{-1})/\Gam$ with a shape parameter
$\Gam=\Om_mh\exp\left[-\Om_b(1+\sqrt{2h}/\Om_m)\right]$ that accounts
for baryonic effects as described in~\Cite{Sugiyama:1994ed}.
For more accurate results, one could use a numerical transfer function,
as obtained by codes like CMBFAST~\cite{Seljak:1996is} or CAMB~\cite{Lewis:1999bs};
the results are not expected to be qualitatively different.

In order to study halos, which form where an extended region of
space has an average overdensity which is above threshold, it is
useful to introduce a filter function $W_R(|\x|)$, and consider the
smoothed density field (around one point, which we take as the
origin),  
\be
\del_R = \int \frac{d^3 k}{(2 \pi)^3} \tW(k R) \del(\k) \, ,
\la{delR}
\ee
where $\tW(k R)$ is the Fourier transform of the filter function. 
For all numerical calculations we will use the spherical top-hat
filter in real space, whose Fourier transform $\tW(kR)$ is given by
\be
\tW(y) = \frac{3}{y^3}\left(\sin y - y\cos y \right)\,.
\la{filter-sharpx}
\ee
This choice allows us to have a well-defined relation between length 
scales and masses, namely $M=(4\pi/3)\Om_m\rho_cR^3$ with
$\rho_c = 3 H_0^2/(8 \pi G) = 2.75\cdot10^{11} h^{-1}\Msol(h^{-1}{\rm
  Mpc})^{-3}$. However it introduces some complexities in the
analysis, which we will comment on later. By using \eqns{delR} and
\eqref{transfer} we have, for the $3$-point function, 
\be
\avg{\del_{R_1} \del_{R_2} \del_{R_3}}_c
= \int \frac{d^3 k_1}{(2 \pi)^3} \frac{d^3 k_2}{(2 \pi)^3} \frac{d^3
  k_3}{(2 \pi)^3} \tW(k_1 R_1) \tW(k_2 R_2) \tW(k_3 R_3) \Cal{M}(k_1)
\Cal{M}(k_2) \Cal{M}(k_3) \avg{\R(\k_1) \R(\k_2) \R(\k_3)}_c \, ,
\ee
where the subscript $c$ denotes the connected part, and analogous
formulae are valid for the higher order correlations.

\subsection{Shapes of non-Gaussianity} 
\noindent
The function $\avg{\R(\k_1) \R(\k_2) \R(\k_3)}_c$ encodes
information about the physics of the inflationary epoch. 
By translational invariance, it is proportional to a
momentum-conserving delta function: 
\be
\avg{\R(\k_1) \R(\k_2) \R(\k_3)}_c = (2 \pi)^3
\del_D(\k_1+\k_2+\k_3) 
B_\R(k_1,k_2,k_3) \, ,
\ee
where the (reduced) bispectrum $B_\R(k_1,k_2,k_3)$ depends only on the
magnitude of the $k$'s by rotational invariance.
According to the particular model of inflation, the
bispectrum will be peaked about a particular shape of the triangle. 
The two most common cases are the squeezed (or local) NG,
peaked on squeezed triangles $k_1 \ll k_2 \simeq k_3$, and the
equilateral NG, peaked on equilateral triangles $k_1
\simeq k_2 \simeq k_3$. Indeed, one can define a scalar product of
bispectra, which describes how sensitive one is to a
NG of a given type if the analysis is performed using
some template form for the bispectrum.
As expected, the local and equilateral shapes are approximately orthogonal
with respect to this scalar product~\cite{Babich:2004gb}.
We will now describe these two models in more detail. 

\vskip 0.15in
\noindent
\underline{\it\large The local model:}
\vskip 0.1in
\noindent
The local bispectrum is produced when the NG is generated 
outside the horizon, for instance in the curvaton
model~\cite{Lyth:2002my, Bartolo:2003jx} or in the inhomogeneous
reheating scenario~\cite{Dvali:2003em}. 
In these models, the curvature perturbation can be written in the
following form,
\be
\R(\x) = \R_g(\x) + \frac{3}{5} \fnl^{\rm loc} \left(\R_g^2(\x) -
\avg{\R_g^2}\right) 
+ \frac{9}{25} \gnl \R_g^3(\x) \, , 
\ee
where $\R_g$ is the linear, Gaussian field. We have included also a
cubic term, which will generate the trispectrum at leading order. 
The bispectrum is given by 
\be
B_\R(k_1,k_2,k_3) = \frac{6}{5} \fnl^{\rm loc} \left[
  P_\R(k_1) P_\R(k_2) + {\rm cycl.} \right] \, ,
\ee
where ``cycl.'' denotes the 2 cyclic permutations of the wavenumbers,
and $P_\R(k)$ is the power spectrum given by $P_\R(k)=Ak^{n_s-4}$.
The trispectrum is given by
\begin{multline}
\avg{\R(\k_1) \R(\k_2) \R(\k_3) \R(\k_4)}_c =
(2 \pi)^3 \del_D(\k_1+\k_2+\k_3+\k_4) \\
\times \left[ \frac{36}{25} \fnl^2 \sum_{\substack{b < c \cr a \neq b,
      c}} P_\R(|\k_a + \k_b|) P_\R(k_b) P_\R(k_c) 
+ \frac{54}{25} \gnl \sum_{a<b<c} P_\R(k_a) P_\R(k_b) P_\R(k_c)
\right] \, . 
\end{multline}

\vskip 0.15in
\noindent
\underline{\it\large The equilateral model:}
\vskip 0.1in
\noindent
Models with derivative interactions of the inflaton
field~\cite{Alishahiha:2004eh, ArkaniHamed:2003uz, Creminelli:2003iq}
give a bispectrum which is peaked around equilateral configurations,
whose specific functional form is model dependent.
Moreover, the form of the bispectrum is usually not convenient to use
in numerical analyses.
This is why, when dealing with equilateral NG,
it is convenient to use the following parametrization, given
in~\Cite{Creminelli:2005hu},
\begin{equation}
B_\R(k_1,k_2,k_3) = \frac{18}{5} \fnl^{\rm equil} A^2
\Big[ \frac{1}{2 k_1^{4-n_s}  k_2^{4-n_s}}
+ \frac{1}{3 (k_1 k_2 k_3)^{2 (4-n_s)/3}} \\
- \frac{1}{(k_1 k_2^2 k_3^3)^{(4-n_s)/3}} + \text{5 perms.} \Big] \, .
\la{equilateral}
\end{equation}
This is peaked on equilateral configurations, and its scalar product
with the bispectra produced by the realistic models cited above is
very close to one. 
Therefore, being a sum of factorizable functions, it is the standard
template used in data analyses. 


\section{Random walks and the halo mass function}
\la{calc-f}
\noindent
We now turn to the main calculation of the paper.
The non-Gaussian halo mass function can be obtained by calculating the
barrier first crossing rate \Cal{F}\ of a random walk with
non-Gaussian noise, in the presence of an absorbing barrier. This can
be done perturbatively, starting from a path integral approach as
prescribed by MR~\cite{Maggiore:2009rv,Maggiore:2009rx} and the mass
function can be shown to be $f(\sig) = 2\sig^2\Cal{F}(\sig)$. As
discussed by MR, the calculation of $f$ involves certain assumptions
regarding the type of filter used and also the location of the
barrier. In particular, the formalism is simplest for a sharp filter
in $k$-space, and using the spherical top-hat of~\eqn{filter-sharpx}
introduces complications in the form of non-Markovian
effects. Further, in order to make the spherical collapse ansatz more
realistic and obtain better agreement with $N$-body simulations, MR
show that it is useful to treat the location of the barrier $\delc$ as
a stochastic variable itself, and allow it to diffuse. For the time
being, we will ignore these complications, and will return to their
effects in section \ref{diffuse-filter}. 

To make the paper self-contained, we begin with a brief review of the
path integral approach to the calculation of the mass function. The
reader is referred to~\Cite{Maggiore:2009rv} for a more pedagogical
introduction. In the path integral approach, one treats the
variance $\sig^2_R\equiv\avg{\hat\del_R^2}$ as a ``time'' parameter,
$t\equiv\sig^2_R$, and considers the random walk followed by the
smoothed density field $\hat\del_R$ as this ``time'' is increased in
discrete steps starting from small values (equivalently, as $R$ is
decreased from very large values). Here $\hat\del_R(\vec{x})$ is a
stochastic quantity in real space due to the stochasticity inherent in
the initial conditions. We use the notation $\hat\del_R$ to
distinguish the stochastic variable from the values it takes, which
will be noted by $\del_i$ below. We probe this stochasticity by
changing the smoothing scale at a fixed location $\vec{x}=0$, thus
making the variable perform a random walk, which obeys a Langevin
equation  
\be
\frac{\p\hat\del}{\p t} = \hat\eta\,,
\la{langevin}
\ee
with a stochastic noise $\hat\eta$ whose statistical properties depend
on the choice of filter used. In particular, for a top hat filter in
$k$-space, the noise is white, i.e. its $2$-point function is a Dirac
delta 
\cite{Bond:1990iw}, 
\be
\avg{\hat\eta(t_1)\hat\eta(t_2)} = \del_{\rm D}(t_1-t_2)
\la{whitenoise}
\ee
The random walk can be described as a trajectory $\{\del_0,\del_1,
\ldots,\del_n\}$ which starts with $\hat\delta$ taking the value
$\del_0=0$ at $t=0$ (or $R\to\infty$ which is the homogeneous limit),
then taking values $\del_i$ at times $t_i$, finally arriving at
$\del_n$ at time $t_n$, with a discrete timestep $\Dt=t_{k+1}-t_k=t_n/n$.
The probability $\Cal{P}(t)$ that the trajectory crosses the barrier
at $\delta_c$ at a time larger than some $t$ (i.e. at scales smaller
than the corresponding $R$ or $M$), is the same as the probability
that the trajectory \emph{did not} cross the barrier at any time
smaller than $t$, so that
\be
\Cal{P}(t) = \int_{-\infty}^{\delc}\rmd \del_1 \ldots \rmd \del_n
W(\{\del_j\};t)\,,
\la{Poft}
\ee
where the probability density over the space of trajectories,
$W(\{\del_j\};t)$ is defined as 
\be
W(\{\del_j\};t) \equiv \avg{\del_{\rm D}(\hat\del(t_1)-\del_1)\ldots
  \del_{\rm D}(\hat\del(t_n)-\del_n)}\, ,
\la{W-def}
\ee  
where $\del_{\rm D}$ is the Dirac delta distribution.
The first crossing \emph{rate} is given by the negative time
derivative of \Cal{P}, $\Cal{F}=-\p_t\Cal{P}$, and the mass function
is then $f=2t\Cal{F}(t)$.
In~\eqn{W-def} one can write the
Dirac deltas using the integral representation $\del_{\rm D}(x) =
\int_{-\infty}^{\infty}{d\lam e^{-i\lam x}/2\pi}$, to obtain
\be
W(\{\del_j\};t) = \int_{-\infty}^{\infty}{\frac{d\lam_1}{2\pi} \ldots
   \frac{d\lam_n}{2\pi} \avg{e^{-i\sum_j\lam_j\hat\del(t_j)}}
  e^{i\sum_j\lam_j\del_j} } \,.
\la{W-lam}
\ee
The object $\avg{e^{-i\sum_j\lam_j\hat\del_j}}$ is the exponential of
the generating functional of the connected Green's functions, and can
be shown to reduce to~\cite{ZinnJustin:2002ru}
\be
\avg{e^{-i\sum_j\lam_j\hat\del_j}} = \exp \left[\sum_{p=2}^\infty
  \frac{(-i)^p}{p!} \sum_{j_1,..,j_p=1}^n\lam_{j_1}\ldots
  \lam_{j_p} \avg{\hat\del_{j_1}\ldots \hat\del_{j_p}}_c  \right]\,,
\la{gen-func}
\ee
where $\avg{\hat\del_{j_1}\ldots \hat\del_{j_p}}_c$ is the connected
$p$-point function of $\hat\del$, with the short-hand notation
$\hat\del_{j} = \hat\del(t_{j})$.

\subsection{Halo mass function: Gaussian case, sharp-$k$ filter}
\la{gaussmark-intro}
\noindent
In the Gaussian case, all connected $n$-point correlators vanish
except for $n=2$, and in the Markovian (sharp-$k$ filter) case which
we are considering, 
the $2$-point function becomes $\avg{\hat\del_j\hat\del_k} = {\rm min}(t_j,t_k)$,
where ${\rm min}(t_j,t_k)$ is the minimum of $t_j$ and $t_k$.
The resulting $n$-dimensional Gaussian integral can be handled in a
straightforward way to obtain 
\be
W^{\rm gm}=\prod_{k=0}^{n-1}{\Psi_{\Dt}(\del_{k+1}-\del_k)} ~;~~~
\Psi_{\Dt}(x) = (2\pi\Dt)^{-1/2} e^{-x^2/(2\Dt)}\,,
\la{Wgm}
\ee
where we follow MR's notation and use the superscript ``gm'' to denote
``Gaussian Markovian''. As MR have shown~\cite{Maggiore:2009rv}, the
resulting expression for $\Cal{P}_{\rm gauss}(t)$ in the continuum
limit $\Dt\to0$ is simply  
\be
\Cal{P}_{\rm gauss} = \int_{-\infty}^{\delc}{d\del_1\ldots d\del_n
  W^{\rm gm}} = \erf{\frac{\nu}{\sqrt{2}}}\,,
\la{Wgmintegral}
\ee
where we use the notation $\nu \equiv \del_c/\sig$. (This in principle
also includes the redshift dependence of the collapse threshold \delc,
see below.) This expression for the continuum
limit probability $\Cal{P}_{\rm gauss}$ is of course a well-known
result going back to Chandrasekhar \cite{Chandrasekhar:1943ws}. This
leads to the standard excursion set result for the Gaussian mass
function $f_{\rm PS}= -2t\p_t|_{\delc}\Cal{P}_{\rm gauss}$, 
\be
f_{\rm PS}(\nu) = \sqrt{\frac{2}{\pi}} \, \nu \, e^{-\nu^2/2}\,,
\la{fgauss}
\ee
where we use the subscript PS (for Press-Schechter), to conform with the
conventional notation for this object.

\subsection{Halo mass function: non-Gaussian case, sharp-$k$ filter} 
\la{calc-f-sharpk}
\noindent
In the non-Gaussian case (but still retaining the sharp-$k$ filter),
the probability density $W(\{\del_j\};t)$ also gets
contributions from connected $n$-point correlators with $n\geq3$,
since these in general do not vanish. These can be handled by
using the relation $\lam_ke^{i\sum_j\lam_j\del_j} = -i\p_k
e^{i\sum_j\lam_j\del_j}$, with $\p_j\equiv\p/\p\del_j$. A
straightforward calculation then shows the mass function to be
\begin{align}
f &= -2 t \left.\frac{\p}{\p  t}\right|_{\delc}
  \int_{-\infty}^{\delc}{\rmd \del_1 \ldots \rmd \del_n
  \exp{\bigg[-\frac{1}{3!}\sum_{j,k,l=1}^n{\avg{\hat\del_j \hat\del_k
          \hat\del_l}_c \p_j\p_k\p_l} }} \nonumber\\
  &\ph{-2 t \frac{\p}{\p t}|_{\delc}
    \int_{-\infty}^{\delc}{d\del_1\ldots d\del_n}} 
  +\frac{1}{4!} \sum_{j,k,l,m=1}^n {\avg{\hat\del_j \hat\del_k
      \hat\del_l \hat\del_m}_c \p_j\p_k\p_l\p_m}  +\ldots  \bigg]
  W^{\rm gm}\,, 
\la{fMR}
\end{align}
where it is understood that one takes the continuum limit $\Dt\to0$
before computing the overall derivative with respect to $t$. 
We will find it useful to change variables from (\delc, $t$) to
($\nu$, $t$), in which case the partial derivative becomes    
\be
- 2 t (\p/\p t)|_{\delc} = \nu (\p/\p\nu)|_t - 2 t (\p/\p t)|_{\nu}
\equiv \nu\p_\nu-2t\p_t\,. 
\la{coordchange}
\ee
It is also useful at this stage to take a small detour and introduce
some notation which we will use throughout the rest of the paper. We
define the scale dependent ``equal time'' functions
\be
\eps_{n-2} \equiv \frac{\avg{\hat\del_R^n}_c}{\sig_R^n} ~~;~~ n\geq3\,,
\la{notes-eq1}
\ee
which as we will see, remain approximately constant over the scales of 
interest. We assume the ordering $\eps_{n-2}\sim\Cal{O}(\ep^{n-2})$ with
$\ep\ll1$, which can be motivated from their origin in inflationary
physics, where one finds $\eps_1\sim\fnl A^{1/2}$,
$\eps_2\sim \gnl A$, etc\footnote{Notationally we distinguish the
  order parameter \ep\ from the specific NG functions
  $\eps_1$ and $\eps_2$.}. Typically we expect $\ep\lesssim10^{-2}$ for
$\fnl\lesssim100$. \fig{fig-eps} shows the behaviour of $\eps_1$ and
$\eps_2$ in the local and equilateral models, as a function of
$t=\sig^2_R$. We see e.g. that $\eps_2$ in the local model is
comparable to $\eps_1^2$.
In the literature one usually encounters the reduced
cumulants $\Cal{S}_n$, which are related to the $\eps_{n-2}$ by
$\eps_1=\sig\Cal{S}_3$, $\eps_2=\sig^2\Cal{S}_4$ and so on. The
motivation for using the $\Cal{S}_n$ comes from the study of
NG induced by nonlinear gravitational effects. However,
as we see from \fig{fig-eps}, when studying \emph{primordial}
NG it is more meaningful to consider the $\eps_n$ which
are approximately scale-independent and perturbatively ordered.
\begin{figure}[t]
\centering
\subfloat[]{\includegraphics[width=0.47\textwidth]{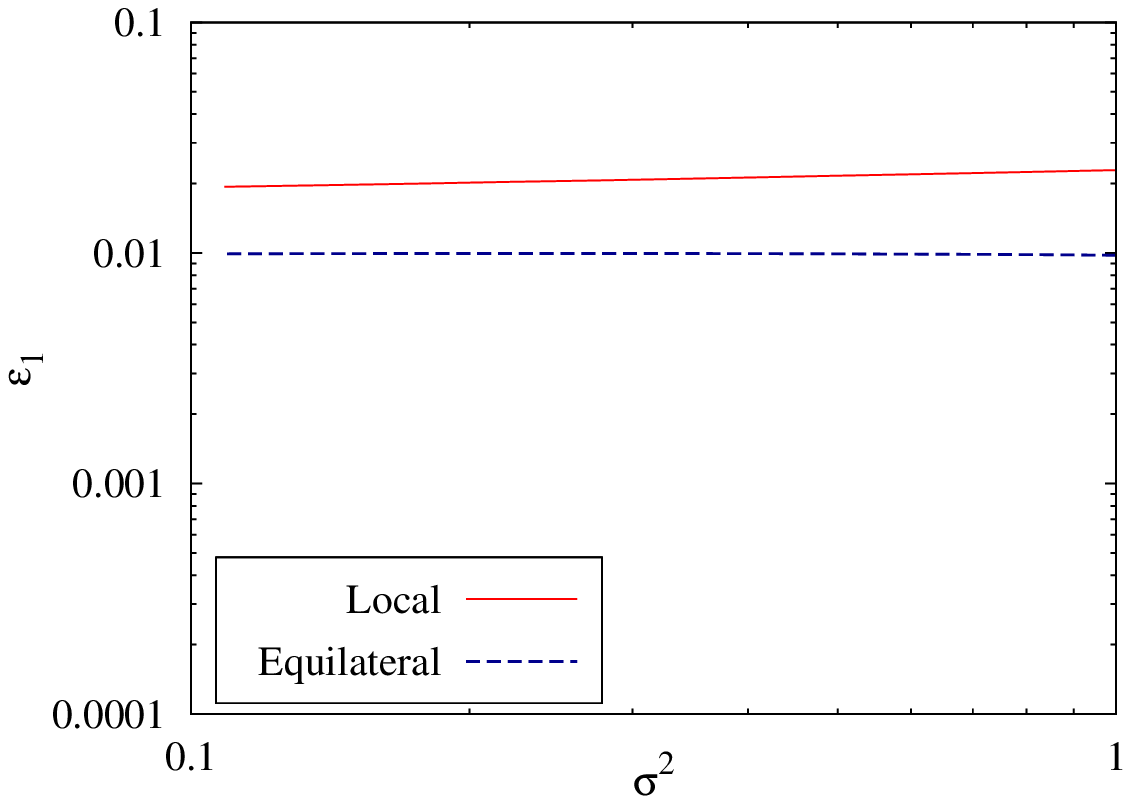}}
\la{fig-eps-a}
\hspace{0.03\textwidth}
\subfloat[]{\includegraphics[width=0.47\textwidth]{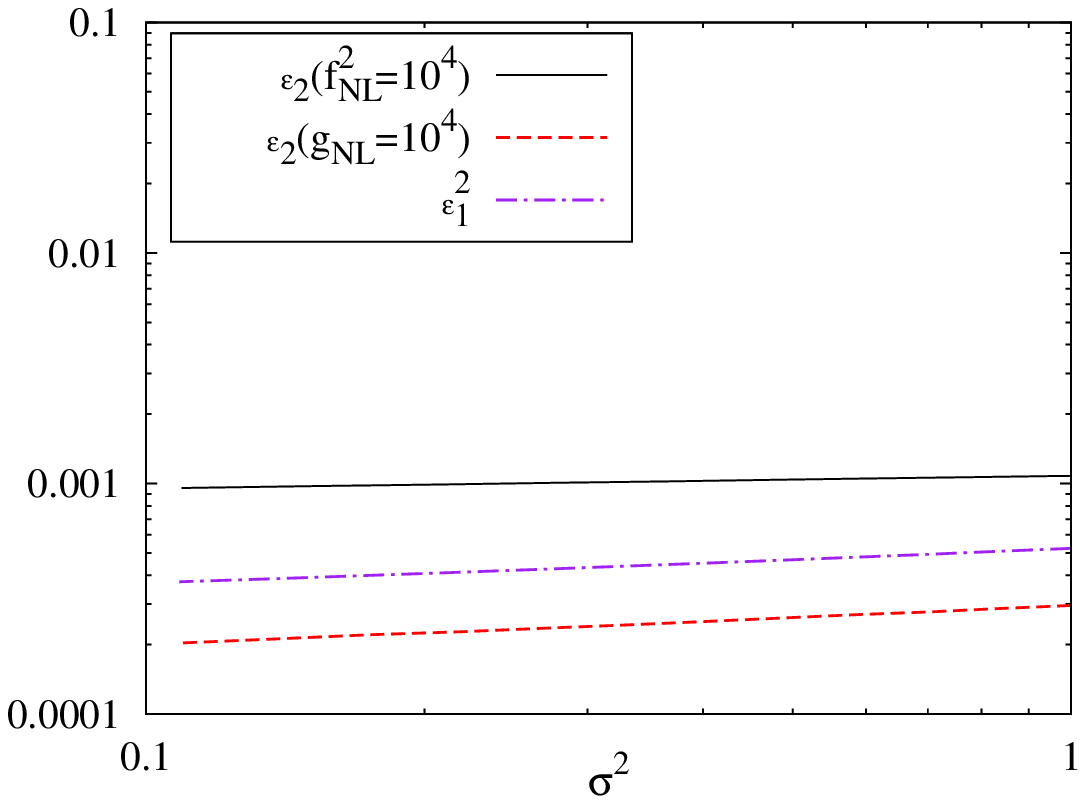}}
\la{fig-eps-b}
\caption{\footnotesize Scale dependence of the $\eps_n$. Panel
  (a) : Behaviour of $\eps_1$ vs. $\sig^2$ in the local and
  equilateral models, for $\fnl=100$ in each case. Panel (b) : 
  Behaviour of $2$nd order ($\sim\ep^2$) terms. We show $\eps_2$ for
  the local model with $\fnl=100$ and $g_{\rm NL}=10^4$. The terms
  proportional to $\fnl^2$ and \gnl\ are shown separately. Also shown
  is $\eps_1^2$ for the same model.}
\la{fig-eps}
\end{figure}

We will soon see that a natural expansion parameter 
that arises in the calculation has the form $\sim\ep\nu$, and we
therefore require that the mass scales under scrutiny are not large
enough to spoil the relation $\ep\nu\ll1$. It turns out that
observationally interesting mass scales can nevertheless be large
enough to satisfy $\ep\nu^3 \sim\Cal{O}(1)$.
\fig{fig-epnu3-epnu} shows the behaviour of $\eps_1\nu^3$ and
$\eps_1\nu$ at different redshifts, as a function of mass, in our
reference \Lam CDM model for local type NG, with
$\fnl^{\rm loc}=100$. The behaviour for the equilateral
NG is qualitatively similar.   
\begin{figure}[t]
\centering
\subfloat[]{\includegraphics[width=0.47\textwidth]{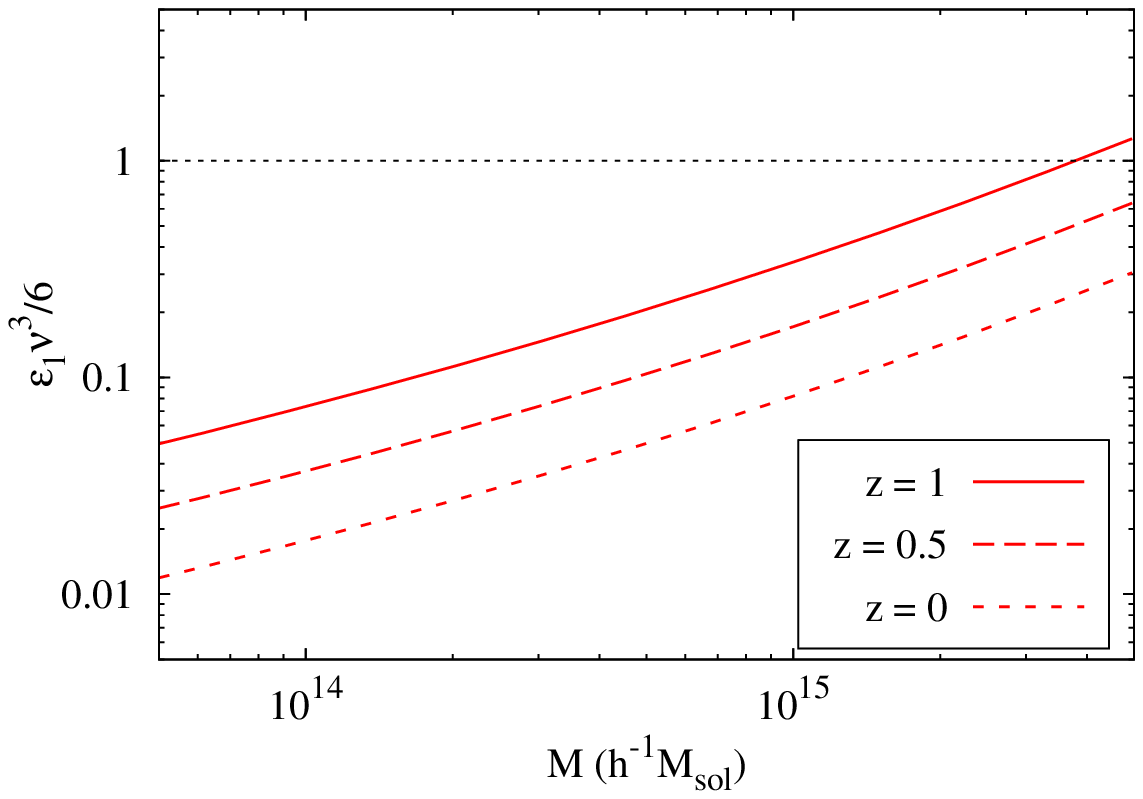}}
\la{fig-epnu3}
\hspace{0.03\textwidth}
\subfloat[]{\includegraphics[width=0.47\textwidth]{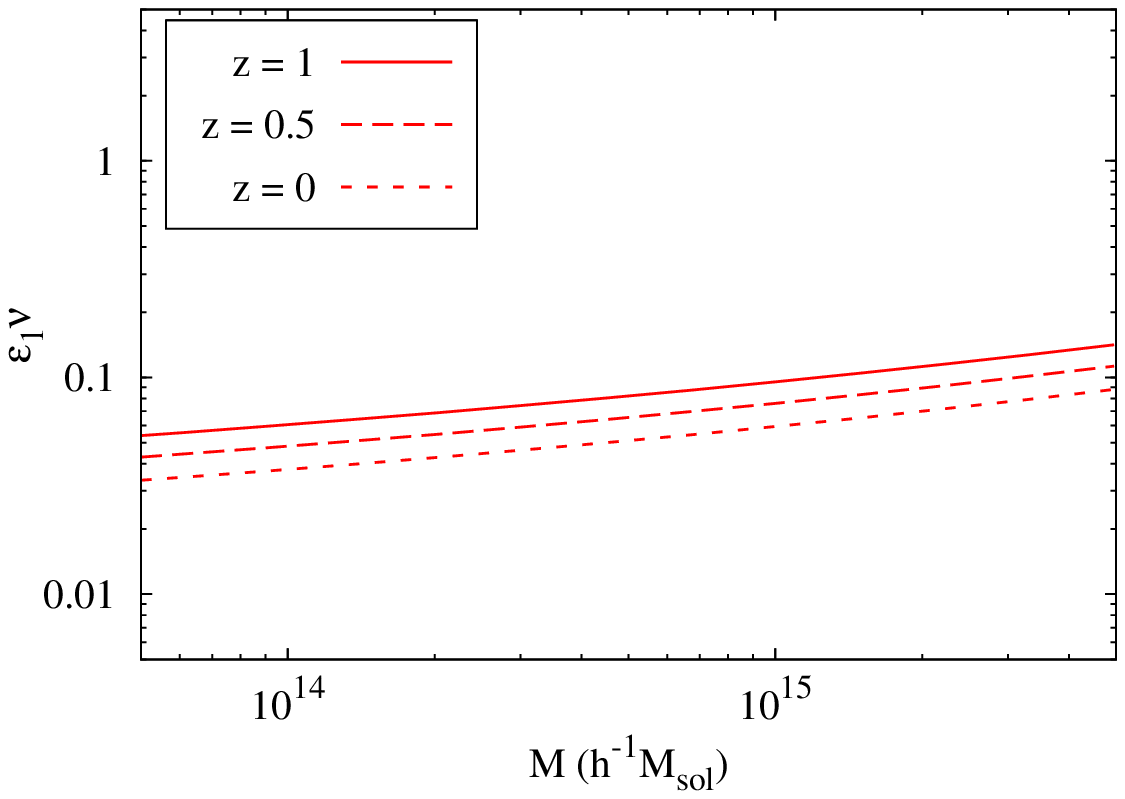}}
\la{fig-epnu}
\caption{\footnotesize Panel (a) : Behaviour of $\eps_1\nu^3/6$
  vs. mass in the local non-Gaussian model, for $\fnl=100$. The three
  curves correspond to different redshifts. The horizontal line
  corresponds to $\eps_1\nu^3/6=1$. Panel (b) : Behaviour of
  $\eps_1\nu$ with the same setup as in panel (a).}
\la{fig-epnu3-epnu}
\end{figure}
The redshift dependence of these quantities comes from the definition
of $\nu$, 
\be
\nu(M,z)\equiv \sqrt{a} \frac{\del_{c0}}{\sig(M)}\frac{D(0)}{D(z)}
\equiv \frac{\delc(z)}{\sig(M)}\,, 
\la{nuofz}
\ee
where we denote the usual spherical collapse threshold as
$\del_{c0}=(3/5)(3\pi/2)^{2/3}\simeq1.686$, reserving \delc\ for the
full, redshift dependent quantity, and $a$ is a parameter accounting for deviations from the simplest collapse model.
In the standard spherical collapse picture we have $a=1$.
A value of $a$ different from unity
(specifically $\sqrt{a}\simeq0.89$) can be motivated by allowing the
collapse threshold to vary stochastically \cite{Maggiore:2009rw}, as
we will discuss in section~\ref{diffuse-filter}. We will soon see that
the object $\ep\nu^3$ appears naturally in the calculation, and to be
definite we will assume $\ep\nu^3\sim\Cal{O}(1)$ for now. In
section~\ref{consistency} we will discuss the effects of relaxing this
condition and probing smaller length scales. 

We now turn to the ``unequal time'' correlators appearing in~\eqn{fMR}.
Since we are concerned with large scales, we are in the
small $t$ limit, and following MR we expand the $n$-point correlators
around the ``final time'' $t$. We can define the Taylor coefficients
\be
\Cal{G}_3^{(p,q,r)} (t) \equiv \left[ \frac{d^p}{dt^p_j}
  \frac{d^q}{d t^q_k} \frac{d^r}{d t^r_l} \avg{\hat\del(t_j)
    \hat\del(t_k) \hat\del(t_l)}_c \right]_{t_j=t_k=t_l=t} \, , 
\la{G3def}
\ee
and then expand
\be
\avg{\hat\del_j \hat\del_k \hat\del_l}_c = \sum_{p,q,r=0}^\infty
\frac{(-1)^{p+q+r}}{p! q! r!} 
\Cal{G}_3^{(p,q,r)} (t) (t-t_j)^p (t-t_k)^q (t-t_l)^r \, .
\la{del3expnn}
\ee
For the $4$-point function we will have an analogous expression
involving coefficients $\Cal{G}_4^{(p,q,r,s)}$. 

Since calculations involving a general set of coefficients
$\Cal{G}_3$, $\Cal{G}_4$, etc. are algebraically rather involved, we
find it useful to first consider an example in which these
coefficients take simple forms. In this toy model we assume that the
$\eps_n$ are exactly constant, and moreover that the $n$-point
correlators take the form\footnote{Throughout the paper we will
  consider at most $4$-point correlators. This truncation is justified
  given our assumptions, as we will see later.}
\be
\avg{\hat\del_j \hat\del_k \hat\del_l}_c=\eps_1 (t_jt_kt_l)^{1/2} ~~;~~
\avg{\hat\del_j \hat\del_k \hat\del_l \hat\del_m}_c=\eps_2 (t_jt_kt_lt_m)^{1/2}
\,. 
\la{toymodel}
\ee
For clarity, we will display details of the calculation only for this
model. We have relegated most of the technical details of our
calculation to Appendix \ref{app-massfunc}. In Appendix
\ref{app-unequaltimetoy} we show that the mass function for this model
can be brought to the form
\begin{align}
f 
&= \left(\frac{2}{\pi}\right)^{1/2} \nu\, e^{-(\eps_1/3!)\p_\nu^3 +
  (\eps_2/4!)\p_\nu^4 +\ldots} \bigg[e^{-\nu^2/2} -
    \frac{1}{4}\eps_1\nu\, e^{-\nu^2/2}+ \frac{5}{16}\eps_1
    \left(\frac{\pi}{2}\right)^{1/2}\erfc{\frac{\nu}{\sqrt{2}}} 
    \nonumber\\
&\ph{\left(\frac{2}{\pi}\right)^{1/2} \nu e^{-(\eps_1/3!)\p_\nu^3 +
  (\eps_2/4!)\p_\nu^4 +\ldots} \bigg[e^{-\nu^2/2}}
    + \frac{1}{8}\left(\eps_1^2-\frac{2}{3}\eps_2\right)
    e^{-\nu^2/2}\left(\nu^2-1\right) + \Cal{O}(\ep^3\nu^3) \bigg]\,,
\la{f-expderiv}
\end{align}
where we ignore terms like $\eps_1\nu\Cal{O}(\nu^{-4})$. The remaining
exponentiated derivative can be handled using a saddle point
approximation. 
We show how to do this in Appendix \ref{app-massfuncsaddlept}. 
The expression for the mass function $f(\nu)$ works out to
\begin{align}
f(\nu) = \left(\frac{2}{\pi}\right)^{1/2} \nu
&\exp{\left[-\frac{1}{2}\nu^2\left(1-\frac{\eps_1}{3}\nu +
    \frac{1}{4}\left(\eps_1^2-\frac{\eps_2}{3}\right)\nu^2 +
  \Cal{O}(\ep^3\nu^3)\right)\right]}\nonumber\\
&\ph{\exp{[-\frac{1}{2}\nu^2()]}}
\times \left(1 - \frac{1}{4}\eps_1\nu\left(3 -
\frac{5}{4\nu^2}\right) + \left(\eps_1^2 -
\frac{\eps_2}{3}\right)\nu^2 + \Cal{O}(\ep^3\nu^3) \right)\,,
\la{f-consteps}
\end{align}
which superficially at least, is comprised of \emph{two} series
expansions, one in the exponential and one as a polynomial, both based
on the small parameter $\ep\nu$ (see however the next section).  

This derivation assumed that $\eps_1$ and $\eps_2$ are constant, and
that the unequal time correlations are given by \eqn{toymodel}. It is
straightforward to relax these assumptions and perform the calculation
with the exact structure of the correlations. Appendix
\ref{app-unequaltimefull} shows how to do this, and the result is
\begin{align}
  f(\nu,t) = &\left(\frac{2}{\pi}\right)^{1/2}\nu\,
\exp\left[-\frac{1}{2}\nu^2\left(1-\frac{\eps_1}{3}\nu +
    \frac{1}{4}\left(\eps_1^2-\frac{\eps_2}{3}\right)\nu^2 +
  \Cal{O}(\ep^3\nu^3)\right)\right]\nonumber\\
&\times
\left \{1-\frac{1}{4} \eps_1\nu
\left(\left(4-c_1\right) 
  +\frac{1}{\nu^2}\left(c_1-\frac{1}{4}c_2-2\right)
  \right)\right. \nonumber\\ 
& \left. \ph{\bigg[\bigg]} + \frac{1}{8}\nu^2\left(
  \eps_1^2\left(11-3c_1\right) -
  2\eps_2\left(1+\frac{1}{3}c_4-\frac{1}{3}\frac{d\ln\eps_2}{d\ln
    t}\right) \right) 
  + \Cal{O}(\ep^3\nu^3) \right\}\,,
\la{f-geneps}
\end{align}
where the functions $c_n(t)$ are defined in \eqn{cndef} and
characterize the behaviour of the unequal time correlations. In our
toy model above, the $c_n$ reduce to unity. Indeed, as a check we see
that the expression in \eqref{f-geneps} reduces to \eqn{f-consteps} if
we take $\eps_1$, $\eps_2$ to be constant and set the $c_n$ to unity. 

One issue which we have ignored so far, is that the definition of
$\nu$ involves the variance $t=\sig^2$ of the \emph{non-Gaussian}
field. Computationally it is more convenient to work with the variance
$\sig_{\rm g}^2$ of the \emph{Gaussian} field in terms of which
cosmological NG are typically defined. We should then
ask whether this difference will require changes in our expressions
for $f$. We start by noting that this difference in variances is of
order $\sim\ep^2$. For example, in the local model one has $\sig^2(R)
= A d_1(R) + A(A\fnl^2)d_2(R)$ where $A\sim10^{-9}$ is an overall
normalization constant, $d_1$ and $d_2$ are scale dependent functions
of comparable magnitude on all relevant scales, and $\ep$ is estimated
as $\ep\sim\fnl A^{1/2}$. We therefore have
$\nu=\delc/\sig=(\delc/\sig_{\rm g})(1+\Cal{O}(\ep^2))$. However, 
with our assumption that $\ep\nu^3\sim\Cal{O}(1)$, we see that this
correction is actually of order
$\sim(\ep^2\nu^2)\nu^{-2}\sim\ep^3\nu^3$, which we have been  
consistently ignoring. We will see that even when we relax the
assumption $\ep\nu^3\sim\Cal{O}(1)$ and probe smaller scales where
$\ep\nu^3 \ll 1$, this correction can still be consistently
ignored. Hence we can safely set $\nu=\delc/\sig_{\rm g}$ in all of
our expressions.

\section{Consistency of the truncation}
\la{consistency}
\subsection{Comparative sizes of terms in the mass function}
\la{compare-fterms}
\noindent
Now that all the derivative operators which we consider important have
been accounted for, we can check whether our final result is
consistently truncated, i.e. whether we have retained \emph{all} terms
at any given order in the expansion. Symbolically, our current result
for the mass function can be written as 
\be
f\sim e^{-\frac12 \nu^2 \left(1 + \ep\nu + \ep^2\nu^2 +
  \Cal{O}(\ep^2, \ep^3\nu^3)\right)} \left[1+ \ep\nu +\frac{\ep}{\nu} +
\ep^2\nu^2 + \Cal{O}(\ep\nu^{-3},\ep^2,\ep^3\nu^3)
\right]\,, 
\la{mf-symbolic-start}
\ee
with the understanding that coefficients are computed (but not
displayed) for all terms except those indicated by the $\Cal{O}()$
symbols. Also, $\ep^2$ refers to both $\eps_1^2$ and $\eps_2$.

Since the expansions involve two parameters, $\ep\nu$ and $\nu^{-2}$,
they make sense only if we additionally prescribe a relation between
these parameters. So far we assumed that \ep\ is fixed and $\nu$ is
such that $\ep\nu^3\simeq1$, which was based on the observation that
the term $\ep\nu^3$ naturally appears in the exponent and is not
restricted in principle to small values. In Appendix
\ref{app-transitionpts} we discuss this condition in more detail, and
also analyse the consequences of relaxing this condition and probing
smaller mass scales. We find that for observationally accessible mass 
scales \emph{larger} than the scale where $\ep\nu^3\simeq\nu^{-3}$,
the single expression  
\be
f\sim e^{- \frac12 \nu^2 \left(1 + \ep\nu + \ep^2\nu^2\right)}  \left[1+
  \ep\nu + \frac{\ep}{\nu} +
  \Cal{O}(\ep^3\nu^5,\ep^2\nu^2,\ep\nu^{-3}) \right]\,, 
\la{mf-symbolic-final}
\ee
is parametrically consistent as it stands -- the terms ignored
are smaller than the smallest terms retained -- and in fact it remains a
very good approximation even when $\ep\nu^3\simeq1$, since the only
``inconsistent'' term then is $\ep\nu^{-1}$, whose effect \emph{reduces}
as $\nu$ increases. On scales where $\ep\nu^3\simeq\nu^{-4}$ and
lower, the theoretical error becomes comparable to or larger than the
quadratic term in the exponential. Plugging back all the coefficients,
we have the following result for the mass function (excluding filter
effects, see section \ref{diffuse-filter}), 
\begin{align}
f(\nu,t) = f_{\rm PS}(\nu)\,
\exp&\left(\frac{1}{6}\eps_1\nu^3 -
    \frac{1}{8}\left(\eps_1^2-\frac{\eps_2}{3}\right)\nu^4
    \right)\nonumber\\ 
&\times
\bigg\{1-\frac{1}{4}\eps_1\nu\left(\left(4-c_1\right)
  +\frac{1}{\nu^2}\left(c_1-\frac{1}{4}c_2-2\right) \right) \nonumber\\
&\ph{\times 1 -
    \frac{1}{4}\eps_1\nu\left(\left(c_1+2-\frac{4}{3}\frac{d\ln\eps_1}{d\ln
    t}\right)\right)}  
+ \Cal{O}(\ep^3\nu^5,\ep^2\nu^2,\ep\nu^{-3}) \bigg\} \,.
\la{f-final}
\end{align}

\subsection{Comparing with previous work}
\la{compare-MR}
\noindent
In this subsection we compare our results with previous work on the
non-Gaussian mass function. As mentioned in the introduction, this
quantity has been computed by several authors in different ways
\cite{Matarrese:2000iz,LoVerde:2007ri,Maggiore:2009rx}. If one considers
the range of validity of the perturbative expansion, the strongest
result so far has been due to MVJ \cite{Matarrese:2000iz}, who
explicitly retain the exponential dependence on $\eps_1$. Their
expression for $f$ can be written as\footnote{The 
  analysis presented by MVJ in fact allows one to retain terms like
  $\sim\ep^2\nu^4$ in the exponential as well, and we have seen that
  when $\ep\nu^3\simeq1$, these terms are as important as the
  polynomial $\ep\nu$ term retained by MVJ. However, since the MVJ
  expression misses unequal time effects of order $\sim\ep\nu$ anyway,
  it is reasonable to compare our results with the expression
  \eqref{f-MVJ}, which is also the one used by most other authors
  (see e.g. \Cites{Giannantonio:2009ak,Carbone:2009zz}).}
\be
f_{\rm MVJ} = f_{\rm PS}(\nu)\frac{e^{\eps_1\nu^3/6}}{
  \left(1-\eps_1\nu/3\right)^{1/2}} 
\left(1-\frac{1}{2}\eps_1\nu\left(1-\frac{2}{3}\frac{d\ln\eps_1}{d\ln
  t} \right)\right)\,. 
\la{f-MVJ}
\ee
The major shortcoming of their result is that it
is based on a Press-Schechter like prescription, and must therefore be
normalized by an appropriate Gaussian mass function, typically taken
to be the one due to Sheth \& Tormen \cite{Sheth:2001dp}. Additionally,
it always misses the contributions due to the unequal time
correlators, which contribute to the terms $\sim\ep\nu$,
$\ep\nu^{-1}$, etc. in \eqn{f-final}. When one considers formal
correctness on the other hand, MR have presented a result based on
explicit path integrals, which accounts for the unequal time
contributions, and which also does not need any \emph{ad hoc}
normalizations (in this context see also Lam \& Sheth
\cite{Lam:2009nb}). Indeed, our calculations in section \ref{calc-f} 
were based on techniques discussed by MR in
\Cites{Maggiore:2009rv,Maggiore:2009rx}. As we discuss
below however, the fact that MR do not explicitly retain the
exponential dependence of $\eps_1\nu^3$, means that their result is
subject to significant constraints on the range of its validity. Their
expression for $f$, ignoring filter effects, is\footnote{We have
  corrected a typo in MR's result \cite{Maggiore:2009rx} : the object
  they define as $\Cal{V}_3$ should appear with an overall positive
  coefficient in their Eqns. (85), (87) and (92).}
\be
f_{\rm MR} = f_{\rm PS}(\nu) \left(1 + \frac{1}{6}\eps_1\nu^3
\left\{ 1-\frac{3}{2\nu^2}\left(4-c_1 \right) -
\frac{3}{2\nu^4}\left(c_1-\frac{1}{4}c_2-2 \right) \right\}  \right)\,.
\la{f-MR-detailed}
\ee
This expression is precisely what one obtains by linearizing our
expression \eqref{f-final} in $\eps_1$, which serves as a check on our
calculations. LMSV \cite{LoVerde:2007ri} present a result based on an
Edgeworth expansion of the type encountered when studying
NG generated by nonlinear gravitational effects
\cite{Bernardeau:2001qr}. The result most often quoted in the
literature is their expression linear in $\eps_1$ (and hence in
$\eps_1\nu^3$), which is 
\be
f_{\rm LMSV,lin} = f_{\rm PS}(\nu)\left(1 +
\frac{1}{6}\eps_1\nu^3\left\{
1-\frac{1}{\nu^2}\left(3-2\frac{d\ln\eps_1}{d\ln t}
\right) - \frac{2}{\nu^4}\frac{d\ln\eps_1}{d\ln t} \right\}\right)\,. 
\la{f-LMSVlin}
\ee
In Appendix B.3 of \Cite{LoVerde:2007ri}, LMSV also give an expression
involving $\eps_1^2$ and $\eps_2$, which can be written as
\begin{align}
f_{\rm LMSV,quad} = f_{\rm PS}(\nu)\bigg[&1 + \frac{1}{6}\eps_1
\left(H_3(\nu) + \frac{2}{\nu}\frac{d\ln\eps_1}{d\ln t}H_2(\nu)
\right) \nonumber\\ 
&+ \frac{1}{72}\eps_1^2\left(H_6(\nu) +
\frac{4}{\nu}\frac{d\ln\eps_1}{d\ln t}H_5(\nu)
\right) + \frac{1}{24}\eps_2\left( H_4(\nu) +
\frac{2}{\nu}\frac{d\ln\eps_2}{d\ln t}H_3(\nu) \right) \bigg]\,, 
\la{f-LMSVquad}
\end{align}
where the $H_n(\nu)$ are the Hermite polynomials of order $n$.
This expression was used by LMSV only as a check on the validity of
their \emph{linear} expression. By comparing with our expression which
is non-perturbative in $\eps_1\nu^3$, we will see below that these
quadratic terms in fact significantly improve LMSV's prediction.

Sticking to the linearized results, we see that the expressions of
both MR and LMSV have the symbolic form 
\be
f\sim e^{-\nu^2/2}\left[1+ \ep\nu^3 + \ep\nu + \frac{\ep}{\nu}
  + \ldots \right]\,,
\la{mf-symbolic-MR}
\ee
where the ellipsis denotes all terms of the type $\ep\nu^{-3}$,
$\ep\nu^{-5}$, etc., as well as all terms containing $\ep^2$. As we
have seen, deciding where to truncate the expression for $f$ is not
trivial, and using our more detailed expression we can ask
whether the expression \eqref{mf-symbolic-MR} is consistent at all the
relevant length scales. Immediately, we see that this expression
cannot be correct once $\ep\nu^3$ becomes close to unity. In Appendix 
\ref{app-truncateothers} we discuss the situation for MR and LMSV at
smaller mass scales.

\section{Effects of the diffusing barrier and the filter}
\la{diffuse-filter}
\noindent
In \Cite{Maggiore:2009rw}, MR showed that the agreement between a
Gaussian mass function calculated using the statistics of random walks,
and mass functions observed in numerical simulations with
Gaussian initial conditions, can be dramatically improved by allowing
the barrier itself to perform a random walk.
This approach is motivated by the fact that the ignorance of the
details of the collapse introduces a scatter in the value of the
collapse threshold for different virialized
objects. The width of this scatter was found by Robertson
\etal\ \cite{Robertson:2008jr} to be a growing function of 
$\sigma(M)$, which is consistent with the physical expectation that
deviations from spherical collapse become relevant at small scales.
The barrier can thus be treated (at least on a first approximation) as 
a stochastic variable whose probability density function obeys a
Fokker-Planck equation with a diffusion coefficient $D_B$, which can
be estimated numerically in a given $N$-body simulation. In
particular, MR found $D_B\simeq0.25$ using the simulations of~\Cite{Robertson:2008jr}.

Conceptually, the variation of the value of the barrier is due to two
types of effects, one intrinsically physical and one more inherent to
the way in which one interprets the results of simulations.
From a physical point of view, the dispersion accounts for deviations
from the simple model of spherical collapse, for instance the effects
of ellipsoidal collapse, baryonic physics, etc. 
On the other hand, the details of the distribution of the barrier (and
therefore the precise value of $D_B$) will depend on the halo finder
algorithm used to identify halos in a particular simulation, since
different halo finders identify collapsed objects with different
properties. MR concluded that the final effect of this barrier
diffusion on large scales can be accounted for in a simple way, by
changing $\del_{c0}\to\sqrt{a}\del_{c0}$ where $a=(1+D_B)^{-1}$.
In practice this change is identical to the one proposed by Sheth
\etal\ \cite{Sheth:1999su}\footnote{A potential issue in this argument 
lies in the assumption of a 
linear Langevin equation for the stochastic barrier $B$, resulting in
a simple Fokker-Planck equation with a constant $D_B$ like the one in
MR, while the distribution of $B$ was found to be approximately
log-normal (and therefore non-Gaussian) in \Cite{Robertson:2008jr}. 
One can see that a Langevin equation of the type $\p_t B = B \xi$ 
(which would produce a log-normal distribution) can be approximated
by $\p_t B = \langle B \rangle \xi$, whenever the fluctuations around
$\langle B \rangle$ are small, and gives a constant diffusion 
coefficient as long as $\langle B \rangle$ is constant. Although
both approximations are reasonable on the scales of interest, non-Gaussian 
and scale dependent corrections to the barrier diffusion should be studied,
since in principle they could be of the same order as the other
corrections retained here. This investigation is left for future work.}. 
As MR argue in section 3.4 of~\Cite{Maggiore:2009rx},
this barrier diffusion effect can also be
accounted for in the \emph{non}-Gaussian case, again by the simple
replacement of $\del_{c0}\to\sqrt{a}\del_{c0}$. It is easy to see that
their arguments go through for all our calculations as well, and we
have implemented this change in our definition of $\nu$ in
\eqn{nuofz}, setting $\sqrt{a}=0.89$. 

In~\Cite{Maggiore:2009rv}, MR also accounted for the non-Markovian
effects of the real space top-hat filter, as opposed to the sharp-$k$ filter
for which the results of section \eqref{calc-f} apply. This is
done by writing the $2$-point function $\avg{\hat\del(R_1) \,
  \hat\del(R_2)}$ calculated using the real space top-hat filter,
as the Markovian value plus a correction, $\avg{\hat\del(R_j) \,\hat\del(R_k)} =
\min(t_j,t_k)+\Delta_{jk}$, and noting that the correction
$\Delta_{jk}$ remains small over the interesting range of length
scales. In fact, MR show that a very good analytical approximation for
the symmetric object $\Delta_{jk}$, is 
\be
\Delta_{jk}\simeq \kappa \min(t_j, t_k) \left(1 - \frac{\min(t_j,
    t_k)}{\max(t_j, t_k)} \right) \, ,
\la{Delta-jk}
\ee
where in our case we find $\kappa(R) \simeq 0.464 + 0.002 R$, with $R$
measured in $h^{-1}$Mpc. The mass function is then obtained by
perturbatively expanding in $\Delta_{ij}$, with the leading effect
being due to the integral 
\be
\int_{-\infty}^{\delc}d\del_1\ldots d\del_n\frac{1}{2} \sum_{j,k=1}^n
\Delta_{jk} \p_j\p_kW^{\rm gm}\,,
\nonumber
\ee
which on evaluation leads to
\be
f_{\rm g, sharp-x}(\nu,t) = \left(\frac{2}{\pi}\right)^{1/2}\nu
\left[(1-\kappa)e^{-\nu^2/2} +
  \frac{\kappa}{2}\Gamma\left(0,\frac{\nu^2}{2}\right) +
  \Cal{O}(\kappa^2) \right] \,, 
\la{f-gaussnonmark}
\ee
where the subscript stands for Gaussian noise with the top-hat filter in real space,
and $\kappa$ introduces a weak explicit $t(=\sig^2)$ dependence. In
\Cite{Maggiore:2009rx} MR proposed an extension of this result to the
non-Gaussian case, by assuming that \emph{all} the non-Gaussian terms
that they computed with the sharp-$k$ filter, would
simply get rescaled by the factor $(1-\kappa)$ at the lowest
order, but otherwise retain their coefficients. Symbolically, their
result (Eqn. 88 of \Cite{Maggiore:2009rx}) is
\be
f_{\rm ng, sharp-x}(\nu,t) \sim \nu
\left[(1-\kappa)e^{-\nu^2/2}\left(1+\ep\nu^3+\ep\nu+ \ep\nu^{-1} \right) +
  \frac{\kappa}{2}\Gamma\left(0,\frac{\nu^2}{2}\right) \right] \,,
\la{f-nongaussnonmark}
\ee
with the specific coefficients of the $\ep\nu^3$, $\ep\nu$ and
$\ep\nu^{-1}$ terms being identical to those in \eqn{f-MR-detailed}.
However, the coefficient of e.g. the $\kappa\ep\nu$ term arises from the
action of an operator $\sim\sum_{j,k}\Delta_{jk}\p_j\p_k$
combining with  the first unequal time operator
$\sim\eps_1t^{1/2}\sum_j(t-t_j)\p_j\sum_{k,l}\p_k\p_l$, and there is
no simple way of predicting its exact value beforehand. Since MR 
explicitly neglect such ``mixed'' terms, their formula is not strictly 
inconsistent, as long as one keeps in mind that the theoretical error
in their expression is of the same order as the terms
$\sim\kappa\ep\nu$ that they include. However, if one wants to
consistently retain such terms, a detailed calculation is
needed\footnote{Notice that this issue is completely decoupled
  from the subtleties in truncation discussed in section
  \ref{consistency} -- this problem remains even at scales where MR's
  expression is formally consistent.}. Our calculations (not
displayed) indicate that the coefficient of the $\kappa\ep\nu$ term 
depends on certain details of the continuum limit of the path
integral near the barrier, which require a more careful study. We are
currently investigating methods of computing these effects. At present
however, we conclude that the mixed terms involving both filter
effects and NG, must be truncated at order $\sim\kappa\ep\nu$.  

Finally, the filter-corrected mass function is also subject to effects
of barrier diffusion. 
Here we make the same assumptions as MR do in~\Cite{Maggiore:2009rw},
namely that the barrier location satisfies a Langevin equation with
white noise and diffusion constant $D_B$, which can be accounted for
by replacing $\kappa\to\ti{\kappa}=\kappa/(1+D_B)=a\kappa$.
However, it is difficult to theoretically predict the unequal time
behaviour of the barrier correlations, and these simple assumptions
must also be tested, 
perhaps by suitably comparing with the detailed results of Robertson
\etal\ \cite{Robertson:2008jr}. We leave this for future work.
Our final expression for the mass function, corrected for effects of
the diffusing barrier and the top-hat real space filter, is
\begin{align}
f(\nu,t) = f_{\rm PS}(\nu)&\bigg(1- \ti{\kappa} +
\Cal{O}(\ti{\kappa}^2) \bigg)
\exp\left[\frac{1}{6}\eps_1\nu^3 -
    \frac{1}{8}\left(\eps_1^2-\frac{\eps_2}{3}\right)\nu^4
    \right]\nonumber\\ 
&\times\bigg\{ 
1+\frac{\left(1-2d\ln\ti{\kappa}/d\ln
  t\right)}{1-\ti{\kappa}}\ti{\kappa}\nu^{-2} 
\left( 1 -  2\nu^{-2}\right) 
-\frac{1}{4}\eps_1\nu\left(\frac{c_1}{1-\ti{\kappa}}
+  4-2c_1 \right)
\nonumber\\
&\ph{\times\bigg\{ 1+}
-\frac{1}{4}\eps_1\nu^{-1}\left(c_1-\frac{1}{4}c_2-2\right)
\nonumber\\ 
&\ph{\times\bigg\{ 1+-1}
 + \Cal{O}(\ti{\kappa}^2\nu^{-2}, \ti{\kappa}\ep\nu,\ti{\kappa}\nu^{-6})    
 + \Cal{O}(\ep^2\nu^2,\ep^3\nu^5,\ep\nu^{-3}) \bigg\} \,,
\la{f-fullcorrected}
\end{align}
where we have chosen to account for the scale independent
$\Cal{O}(\ti{\kappa}^2)$ error arising from filter
effects, as an overall normalization uncertainty, and have explicitly
displayed the orders of the various terms we ignore. Here $f_{\rm
  PS}(\nu)$ is given by \eqn{fgauss} with $\nu(M,z)$ defined in
\eqn{nuofz}.  

To summarize, \eqn{f-fullcorrected} gives an analytical expression for
the non-Gaussian mass function.
This expression is based on approximations that are valid over a larger range of length scales
than the ones presented by MR and LMSV, and incorporates effects which
are ignored in the expression presented by MVJ and LMSV.
Like all these other mass functions, it suffers from the errors introduced by filter effects.
However, the largest of these can be
accounted for as an overall normalization constant, which can be fixed
using, for instance, results of a Gaussian simulation. Among the NG
functions $\eps_1$, $\eps_2$ and the $c_n$ which appear in the mass
function, the most important ones over the mass range of interest are
$\eps_1$, $\eps_2$ and $c_1$ which are nearly constant. In principle
though, all these functions must be computed numerically
for every mass scale of interest, and indeed all the plots in this
paper use the results of such numerical calculations. However, since
this is somewhat tedious to do in practice, in \tab{fits} we provide
analytical approximations for $\eps_1$, $\eps_2$, $c_1$, $c_2$ and
$c_3$, for the local and equilateral case as a function of
$\sig^2$. As mentioned earlier, all these quantities are independent
of redshift, although they depend on the choice of cosmological
parameters in a complicated way in general due to the presence of the
transfer function in their definitions. However, the dependence on
$\sig_8$ is simple, and one can check that we have
$\eps_1\propto\sig_8$, $\eps_2\propto\sig_8^2$ and that the $c_n$ are
independent of $\sig_8$. Recall that the $c_n$ are also independent of
\fnl\ and \gnl.  For completeness, in \tab{fits} we also give
approximations for the filter parameters $\ti{\kappa}$ and
$d\ln\ti{\kappa}/d\ln t$ which appear in the mass function, as
functions of $\sig^2$.   
\begin{table}[t]
\centering
\small
\begin{tabular}{|c|c|c|c|}
\hline 
Parameter & \multicolumn{3}{|c|}{Fitting form $b+c\, t^n$} \\ [0.25ex]\hline
\ph{equ}Local NG\ph{ila} & $b$ & $c$ & $n$  \\ [0.5ex]\hline 
$\eps_1$ & $0.0096$ & $0.015$ & $0.18$ \\ [0.5ex]\hline
$c_1$ & $0.98$ & $0.073$ & $0.094$ \\ [0.5ex]\hline
$c_2$ & $3.15$ & $0.79$ & $0.69$ \\ [0.5ex]\hline
$c_3$ & $2.15$ & $0.45$ & $0.65$ \\ [0.5ex]\hline
\raisebox{-0.25ex}{$\eps_2 (\fnl^2)$} & $-0.0049$ & $0.0059$ & $0.011$ \\ [0.5ex]\hline
$\eps_2 (\gnl)$ & $7.9\cdot10^{-4}$ & $0.0022$ & $0.25$ \\ [0.5ex]\hline
\multicolumn{4}{c}{}\\[0.25ex]
\end{tabular}
\hspace{0.1in}
\begin{tabular}{|c|c|c|c|}
\hline 
Parameter & \multicolumn{3}{|c|}{Fitting form $b+c\, t^n$} \\ [0.25ex]\hline
Equilateral NG & $b$ & $c$ & $n$  \\ [0.5ex]\hline 
$\eps_1$ & $0.01$ & $-4\cdot10^{-4}$ & $1.25$ \\ [0.5ex]\hline
$c_1$ & $1.03$ & $-0.052$ & $0.30$ \\ [0.5ex]\hline
$c_2$ & $2.32$ & $0.93$ & $0.49$ \\ [0.5ex]\hline
$c_3$ & $1.72$ & $0.36$ & $0.54$ \\ [0.5ex]\hline\hline
Filter & \multicolumn{3}{|c|}{} \\ [0.25ex]\hline
$\ti{\kappa}$ & $0.36$ & $0.015$ & $-0.47$ \\ [0.5ex]\hline
\raisebox{-0.35ex}{$d\ln\ti{\kappa}/d\ln t$} & $0.046$ & $-0.064$ &
$-0.17$ \\ [0.5ex]\hline 
\end{tabular}
\caption{\footnotesize Analytical approximations for the various NG
parameters, in the local and equilateral cases,
as a function of $t=\sig^2$, in the range $2\cdot10^{13}<M/(h^{-1}\Msol)
<5\cdot10^{15}$, for $\fnl=100$ and $\gnl=10^4$. We have
$\eps_1\propto\fnl$ in both cases, and for $\eps_2$ in the local case
we give separate approximations for the terms proportional to $\fnl^2$
and \gnl. We do not consider $\eps_2$ in the equilateral case, since
the trispectrum in this case is highly model dependent. We also give
approximations for the filter parameters $\ti{\kappa}$ and
$d\ln\ti{\kappa}/d\ln t$ as functions of $t$, in the same mass
range. The errors on all the approximations are less than $1\%$,
except for $\eps_2(\fnl^2)$ where the error is $\sim6\%$. This was due
to numerical difficulties in calculating this object. These
approximations of course depend on our choice of cosmological
parameters.}    
\la{fits}
\end{table}

\section{Results and Discussion}
\la{results}
\noindent
In this section we conclude with our final results for the
non-Gaussian halo mass function, comparing our approach with previous
work. In principle, we should compare the full expressions for the
mass functions of various authors with ours. However, recall that for
MVJ and LMSV one has to multiply an analytically predicted ratio
$R_{\rm ng} = f(\nu,M,\fnl)/f(\nu,M,\fnl=0)$ with a suitable
\emph{Gaussian} mass function based on fits to simulations. 
It is not clear how to compute theoretical error bars on the latter.
On the other hand, the object $R_{\rm ng}$ itself is an unambiguous
theoretical prediction of every approach, that is MVJ, LMSV, MR and
our work, and we can compute theoretical errors on it. In this work,
we will restrict ourselves to comparing the different expressions for
$R_{\rm ng}$. In future work, we hope to compare both $R_{\rm ng}$ and
the full mass function with the results of $N$-body simulations. 

\begin{figure}[t]
\centering
\includegraphics[height=0.35\textheight]{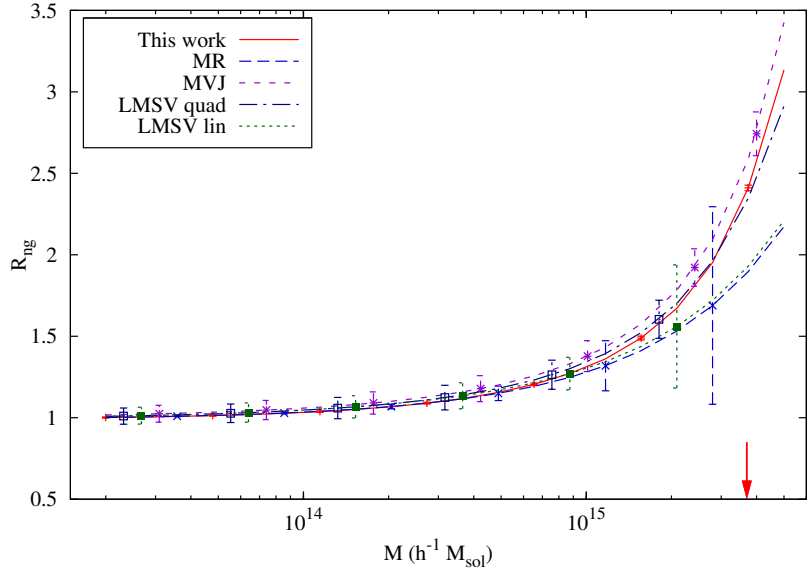}
\caption{\footnotesize Theoretical comparison of the different mass
  functions at $z=1$, without the filter effects, i.e. setting
  $\ti{\kappa}=0$. We plot the ratio $R_{\rm ng}$ of the non-Gaussian
  and Gaussian mass functions, in the local model with $\fnl=100$ and
  $\gnl=0$. See main text for a discussion of the error bars. The
  arrow indicates the mass scale where $\eps_1 \nu^3/6 = 1$,
  i.e. where the expansions of LMSV (both linearized and quadratic)
  and MR break down.}  
\label{final-comparison-z1}
\end{figure}

\begin{figure}[t]
\centering
\includegraphics[height=0.35\textheight]{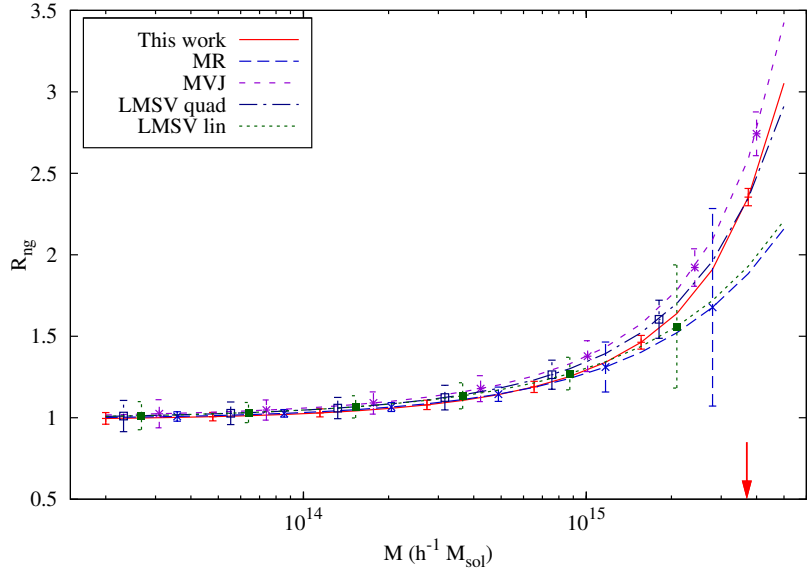}
\caption{\footnotesize Same as \fig{final-comparison-z1}, but
  including filter effects. These affect only the error bars for MVJ
  and LMSV, and they affect both the curve and the error bars for MR and
  our result. For MR and our result, the Gaussian mass function used
  to construct the ratio $R_{\rm ng}$, is taken as the non-Gaussian
  result at $\fnl=0$, and hence includes filter effects.} 
\label{final-comparison-z1k}
\end{figure}

In~\fig{final-comparison-z1} and~\fig{final-comparison-z1k} we
plot the ratio $R_{\rm ng}$, respectively without and with the filter
effects, at redshift $z=1$. In this way we can explicitly disentangle
the errors due to an approximate treatment of non-Gaussian effects,
from those due to the filter effects. We compare our expression
\eqref{f-fullcorrected} with the expressions of
MR~\eqref{f-MR-detailed}, LMSV~\eqref{f-LMSVlin} and 
\eqref{f-LMSVquad}, and of MVJ~\eqref{f-MVJ}. Notice that, when
considering the filter effects, the Gaussian function that enters in
the ratio $R_{\rm ng}$ is defined to be the function with $\fnl=0$
(i.e. without NG but with filter effects when present). We use the
local model, setting $\fnl=100$ and $\gnl=0$, and use the reference
\Lam CDM cosmology described in section \ref{models}.  
We do not explicitly show the final results for the equilateral model,
but they are qualitatively similar. As is commonly done in the
literature, we modify the LMSV and MVJ curves by applying the
Sheth \etal\ correction of $\del_{c0}\to\sqrt{a}\del_{c0}$. 
An identical correction is already present in the expressions
\eqref{f-fullcorrected} and \eqref{f-MR-detailed} due to the barrier
diffusion. We set $\sqrt{a}=0.89$, which is the value inferred by MR
in \Cite{Maggiore:2009rw} using the simulations of Robertson
\etal~\cite{Robertson:2008jr}. We wish to emphasize a feature that our
calculation shares with that of MR, which is that the constant $a$ is
the only parameter whose value depends on the output of $N$-body
simulations. The rest of the calculation for the mass function is
completely analytical and from first principles.

To make the comparison meaningful, we introduce theoretical error bars
on the curves. These error bars have no intrinsic statistical meaning
-- they simply keep track of the absolute magnitude of the terms that
are ignored in any given prescription for computing the mass function.
As we have discussed at length in section \ref{consistency}, these
theoretical errors are scale dependent.
The estimated error magnitude for each point is the maximum among the 
terms ignored in the expression.
More explicitly, the errors for the linearized LMSV
expression~\eqref{f-LMSVlin} are estimated as the maximum of
$(\ep\nu^3)^2$ which comes from the expansion of the exponential,  
$\ep\nu$ which is the order of the largest unequal time terms missing,
and $\tilde{\kappa}\nu^{-2}$ which comes from the filter effects.
The errors for the LMSV expression~\eqref{f-LMSVquad} are similarly
estimated as the maximum of $(\ep\nu^3)^3$, $\ep\nu$ and
$\tilde{\kappa}\nu^{-2}$.  
The largest error for the MVJ expression~\eqref{f-MVJ} is the maximum
of $\ep\nu$ (unequal time terms) and $\tilde{\kappa}\nu^{-2}$ (filter
effects). Finally, the error for the MR
expression~\eqref{f-MR-detailed} is the maximum of $(\ep\nu^3)^2$
from the expansion of the exponential, 
$\ep \nu^{-3}$ from the largest
unequal time terms ignored, and $\tilde{\kappa}^2 \nu^{-2}$ and
$\tilde{\kappa} \ep \nu$ from the filter effects. We include the
filter effects and the associated errors only
in~\fig{final-comparison-z1k}.

From these figures, we can draw some interesting conclusions.
First of all, we see that it is important to retain terms which are
quadratic in the NG, either with a saddle point method
like in MVJ and in our formula, or by expanding the exponential up to
second order, like in LMSV. 
Actually, we argue that it is correct to keep the exponential,
otherwise the expansion breaks down when $\ep \nu^3$ is of order unity.
We notice in passing that the term proportional to $\eps_2$ which
comes from the trispectrum, partially cancels with the $\eps_1^2$
term. Secondly, comparing our expression with MVJ's, we can observe
that keeping the unequal time terms allows us to 
sensibly reduce the theoretical errors due to the approximate
treatment of NG. In fact, if these terms are missing,
they provide the largest theoretical error on large scales. 
Instead, the largest theoretical error on small scales comes from the
approximations involved in dealing with a real space top-hat filter,
as is apparent from~\fig{final-comparison-z1k}.

To conclude, in this work we have calculated the non-Gaussian halo
mass function in the excursion set framework, improving over previous
calculations. We started from a path integral 
formulation of the random walk of the smoothed density field,
following~\Cite{Maggiore:2009rv}. 
This allows us to take into account effects due to multi-scale
correlations of the smoothed density field (``unequal time''
correlations), and due to the real space top-hat
filter, which generates non-Markovianities in the random walk. 
We recognize two small parameters in which we perturb: $\ep$, defined
below \eqn{notes-eq1}, 
which measures the magnitude of the primordial NG;
and $\nu^{-1} = \sigma_R/\del_c$, which is small on very large scales.
In order to do a consistent expansion and to estimate the theoretical
errors, one must study the (scale dependent) relation between these
two parameters, which we have discussed in Sec.~\ref{consistency}. 
We then used saddle point techniques which allowed us to
non-perturbatively retain the dependence on $\ep \nu^3$, which
naturally appears in the calculation and whose magnitude becomes of
order unity at high masses and high redshift. 
Finally, we included effects due to the choice of filter function and
due to deviations from spherical collapse, as explained in
Sec.~\ref{diffuse-filter}. 
Our final result is presented in~\eqn{f-fullcorrected}, which we
reproduce here: 
\begin{align}
  f(\nu,t) = f_{\rm PS}(\nu)&\bigg(1- \ti{\kappa} +
\Cal{O}(\ti{\kappa}^2) \bigg)
\exp\left[\frac{1}{6}\eps_1\nu^3 -
    \frac{1}{8}\left(\eps_1^2-\frac{\eps_2}{3}\right)\nu^4
    \right]\nonumber\\ 
&\times\bigg\{ 
1 + \frac{\left(1-2d\ln\ti{\kappa}/d\ln
  t\right)}{1-\ti{\kappa}}\ti{\kappa} 
\nu^{-2}
\left( 1 -  2\nu^{-2}\right) 
-\frac{1}{4}\eps_1\nu\left(\frac{c_1}{1-\ti{\kappa}}
+  4-2c_1 \right)
\nonumber\\
&\ph{\times\bigg\{ 1+}
-\frac{1}{4}\eps_1\nu^{-1}\left(c_1-\frac{1}{4}c_2-2\right)
\nonumber\\ 
&\ph{\times\bigg\{ 1+-1}
 + \Cal{O}(\ti{\kappa}^2\nu^{-2}, \ti{\kappa}\ep\nu,\ti{\kappa}\nu^{-6})    
 + \Cal{O}(\ep^2\nu^2,\ep^3\nu^5,\ep\nu^{-3}) \bigg\} \, .
\la{f-fullcorrected-repeat}
\end{align}
In \tab{fits} we provide analytical approximations for the various
parameters that appear in this expression, which in general must be
computed numerically. We also considered other expressions for the
mass function found in the literature, which use different expansion
methods but do not estimate the theoretical errors. We estimated the
theoretical errors for each formula, and we show comparative plots in 
\fig{final-comparison-z1} and \fig{final-comparison-z1k}. In our work
we have improved over the calculations of MVJ \cite{Matarrese:2000iz}
and LMSV \cite{LoVerde:2007ri} (who ignore unequal time correlations)
and of MR \cite{Maggiore:2009rx} (who do not retain the exponential
dependence on $\ep\nu^3$). We have also demonstrated that the
(linearized) result of LMSV can be significantly improved by retaining
the quadratic terms of their calculation which are usually ignored in
the literature. We find that at large scales and high redshifts, the
biggest theoretical errors are introduced by ignoring the exponential
dependence on $\ep\nu^3$, followed by the neglect of unequal time
correlations. The errors on our expression
\eqref{f-fullcorrected-repeat} are therefore significantly smaller than
those of the others. The strength of our approach lies in the
combination of path integral methods as laid out by MR
\cite{Maggiore:2009rx}, and the saddle point approximation as used by
MVJ \cite{Matarrese:2000iz}. 

Our work can be continued in several directions. First, a thorough
calculation of the effects due to the choice of the filter should be
performed, since they lead to significant uncertainties in our final
expression. This would include a study of the details of the continuum
limit of the path integral near the barrier, and also a study of the
statistics of the barrier diffusion process in the presence of filter
effects. Second, a comparison with $N$-body simulations should be
performed, in order to quantitatively assess the possibility of
constraining NG using our work. Third, it would be interesting to
study how to account for the effects of ellipsoidal collapse, in a
framewrok such as the one employed in this paper.  Finally, an
application to the void statistics along the same lines should be
feasible. The problem here is made more interesting by the presence of
\emph{two} barriers, as discussed by Sheth \& van de Weygaert
\cite{Sheth:2003py}, and since voids probe larger length/mass scales
than halos, they constitute a promising future probe of primordial NG
\cite{Kamionkowski:2008sr}.

\section*{Acknowledgements}
It is a pleasure to thank Stefano Borgani, Paolo Creminelli, Francesco
Pace, Emiliano Sefusatti, Ravi Sheth, Licia Verde and Filippo Vernizzi for useful
discussions. 

\section*{Appendix}
\appendix
\numberwithin{equation}{section}
\section{Mass function calculation}
\la{app-massfunc}
\subsection{Equal time vs. unequal time terms}
\la{app-unequaltimetoy}
In this appendix we show how the exponentiated derivative operator in
the path integral can be handled by separating the contributions of
the equal time and unequal time correlations. While this calculation
assumes the toy model introduced in \eqn{toymodel}, it easily
generalizes to the more realistic case as we discuss later.

Using the first few terms of the unequal time expansions, in our toy
model one can write  
\begin{align}
&\sum_{j,k,l=1}^n{\avg{\hat\del_j \hat\del_k \hat\del_l}_c\,\p_j\p_k\p_l} =
\eps_1t^{3/2}\bigg(\sum_{j,k,l=1}^n{\p_j\p_k\p_l}
-\frac{3}{2}\sum_{j=1}^n{(1-\frac{t_j}{t})\p_j}\sum_{k,l=1}^n{\p_k\p_l} 
\nonumber\\ 
&\ph{\sum_{j,k,l=1}^n{\avg{\hat\del_j \hat\del_k \hat\del_l}\p_j\p_k\p_l} = \eps_1}
- \frac{3}{8}\sum_{j=1}^n{(1-\frac{t_j}{t})^2\p_j}  
\sum_{k,l=1}^n{\p_k\p_l}
+\frac{3}{4}\sum_{j,k=1}^n{(1-\frac{t_j}{t})(1-\frac{t_k}{t})\p_j\p_k}
\sum_{l=1}^n{\p_l}+\ldots \bigg)\,,
\la{3pt}\\ 
&\sum_{j,k,l,m=1}^n{\avg{\hat\del_j \hat\del_k \hat\del_l
    \hat\del_m}_c\,\p_j\p_k\p_l\p_m} 
= \eps_2t^2 \bigg(\sum_{j,k,l,m=1}^n{\p_j\p_k\p_l\p_m} -
2\sum_{j=1}^n{(1-\frac{t_j}{t})\p_j}\sum_{k,l,m=1}^n{\p_k\p_l\p_m}
+ \ldots \bigg)\,.
\la{4pt}
\end{align}
These derivative operators are exponentiated in the path integral, and
act on $W^{\rm gm}$. One simplification that occurs in our toy model,
is that the path integral in~\eqn{fMR} becomes a function only of
$\nu$ (although this is not obvious at this stage), and hence
eventually only the $\nu\p_\nu$ part of the overall derivative
contributes. However, the structure of the exponentiated derivatives
is still rather formidable. Moreover, the truncation of the series at
this stage is based more on the intuition that higher order terms
should somehow be smaller, rather than on a strict identification of
the small parameters. In fact, we will see in detail
in section \ref{consistency} that the issue of truncation involves
several subtleties. 

To make progress, it helps to analyze the effect on $W^{\rm gm}$ of
each of the terms in the above series, \emph{before} exponentiation.
The leading term in~\eqn{fMR} involves the multiple integral of
$W^{\rm gm}$, which is just the quantity $\Cal{P}_{\rm gauss}$
encountered in \eqn{Wgmintegral}. The operator $\nu\p_\nu$
acts on the error function to give the Gaussian rate of \eqn{fgauss}.  
Next, notice that the action of the operator $\sum_{j=1}^n{\p_j}$ on
\emph{any} function $g(\del_1,\ldots \del_n)$ under the multiple
integral, is simply 
\be
\int_{-\infty}^{\delc}{d\del_1\ldots d\del_n\sum_{j=1}^n{\p_j}g} =
\frac{\p}{\p\delc} \int_{-\infty}^{\delc}{d\del_1\ldots d\del_ng}\,,
\la{delcderivative}
\ee
Using this, and the fact that $t^{1/2}(\p/\p\delc)|_t = \p_\nu|_t$, we
see that the leading term in \eqn{3pt} (i.e. the term with no
powers of $(1-t_j/t)$), leads to a term involving 
\be
\eps_1\nu\p_\nu(\p_\nu)^3\erf{\nu/\sqrt{2}}
\sim f_{\rm PS}\eps_1\nu^3(1+\Cal{O}(\nu^{-2}))\,, 
\nonumber
\ee
The problem with this term is that the quantity $\eps_1\nu^3$ can be of
order unity, and hence cannot be treated perturbatively.
To be consistent, we should keep all terms involving powers of
$\eps_1\nu^3$. Luckily, this can be done in a straightforward way due
to the result in \eqn{delcderivative}. We see that the entire
exponential operator
$\exp[-(\eps_1t^{3/2}/3!)\sum_{j,k,l=1}^n{\p_j\p_k\p_l}]$ in  
\eqn{fMR} can be pulled across the multiple integral and converted to
$\exp{[-(\eps_1/3!)\p_\nu^3]}$ acting on the remaining integral. 
Similarly, the operator
$\exp{[(\eps_2t^2/4!)\sum_{j,k,l,m=1}^n{\p_j\p_k\p_l\p_m}]}$ can be
pulled out and converted to $\exp{[(\eps_2/4!)\p_\nu^4]}$, and the
same applies for all such equal time operators. We will see later that
the action of these operators can be easily accounted for, using a
saddle-point approximation. To summarize, the function $f$ at this
stage is given by 
\begin{align}
f = \nu\, e^{-(\eps_1/3!)\p_\nu^3+(\eps_2/4!)\p_\nu^4+\ldots}\p_\nu
&\int_{-\infty}^{\delc}d\del_1\ldots d\del_n
  \exp\bigg[\frac{1}{3!}\eps_1t^{3/2}\bigg( \frac{3}{2}
      \sum_{j=1}^n{(1-\frac{t_j}{t})\p_j}\sum_{k,l=1}^n{\p_k\p_l}  
   \nonumber\\
  &\ph{\int} + 
   \frac{3}{8}
   \sum_{j=1}^n{(1-\frac{t_j}{t})^2\p_j}\sum_{k,l=1}^n{\p_k\p_l}  
  -\frac{3}{4}
  \sum_{j,k=1}^n{(1-\frac{t_j}{t})(1-\frac{t_k}{t})\p_j\p_k} 
  \sum_{l=1}^n{\p_l}+\ldots\bigg) \nonumber\\
  &\ph{\int}  -\frac{1}{4!}\eps_2t^2\bigg( 2
  \sum_{j=1}^n{(1-\frac{t_j}{t})\p_j}\sum_{k,l,m=1}^n{\p_k\p_l\p_m}+ 
  \ldots \bigg) \bigg]W^{\rm gm}\,.
\la{f-inter}
\end{align}
Now consider the action of the individual terms in the remaining
exponential under the integrals, but without exponentiation. 
From MR \cite{Maggiore:2009rx}, we have the following
results\footnote{The terms in \eqns{notes-eq12a}, \eqref{notes-eq12b} 
  and \eqref{notes-eq12c} are, upto prefactors, the integrals of what
  MR denote as $\Pi^{(3,{\rm NL})}$, $\Pi^{(3,{\rm NNLa})}$ and
  $\Pi^{(3,{\rm NNLb})}$ respectively in \Cite{Maggiore:2009rx}.}, 
\begin{subequations}
\begin{align}
\sum_{j=1}^n{(1 - \frac{t_j}{t})} \sum_{k,l=1}^n{
  \int_{-\infty}^{\delc}{d\del_1\ldots d\del_n\p_j\p_k\p_lW^{\rm gm}}}
&= \left(\frac{2}{\pi}\right)^{1/2}\frac{1}{t^{3/2}}e^{-\nu^2/2} \,, 
\la{notes-eq12a}\\
\sum_{j=1}^n{(1 - \frac{t_j}{t})^2} \sum_{k,l=1}^n{
  \int_{-\infty}^{\delc}{d\del_1\ldots d\del_n\p_j\p_k\p_lW^{\rm gm}}}
&= \left(\frac{2}{\pi}\right)^{1/2}\frac{3}{t^{3/2}}h(\nu)\,,
\la{notes-eq12b}\\
\sum_{j,k=1}^n{(1-\frac{t_j}{t})(1-\frac{t_k}{t})}
\sum_{l=1}^n{\int_{-\infty}^{\delc}{d\del_1\ldots   
    d\del_n\p_j\p_k\p_lW^{\rm gm}}} &=
\left(\frac{2}{\pi}\right)^{1/2}\frac{4}{t^{3/2}}h(\nu)\,,
\la{notes-eq12c}
\end{align}
\end{subequations}
where we have defined
\be
h(\nu)\equiv e^{-\nu^2/2}-\left(\frac{\pi}{2}\right)^{1/2}
\nu\,\erfc{\frac{\nu}{\sqrt{2}}} = \frac{\nu}{2^{3/2}}
\Gam\left(-\frac{1}{2},\frac{\nu^2}{2} \right)\,,
\la{notes-eq13}
\ee
where $\Gam(-1/2,\nu^2/2)$ is an incomplete gamma function.
Let us focus on the term in \eqn{notes-eq12a}. If we linearize in
$\eps_1$ in \eqn{f-inter}, then this term appears with
$\eps_1t^{3/2}\p_\nu$ acting on it, leading to $\sim f_{\rm
  PS}\eps_1\nu\ll f_{\rm PS}$. This term can therefore be
treated perturbatively. Similarly, one can check that the terms given
by \eqns{notes-eq12b} and \eqref{notes-eq12c} also lead to
perturbatively small quantities, which are in fact further suppressed
compared to $\eps_1\nu$ by powers of $\nu^{-2}$. Specifically, one
obtains terms involving $\eps_1\erfc{\nu/\sqrt{2}}$ which, for large
$\nu$, reduces to $\sim f_{\rm
  PS}\cdot\eps_1\nu\cdot\nu^{-2}(1+\Cal{O}(\nu^{-2}))$.  

A few comments are in order at this stage. First, this ordering in
powers of $\nu^{-2}$ is a generic feature of integrals involving an
increasing number of powers of $(1-t_j/t)$ being summed. This can be
understood in a simple way from the asymptotic properties of the
incomplete gamma function, as we show in Appendix
\ref{append-incompgamma}.  
We are therefore justified in truncating the Taylor expansion of the
unequal time correlators, even though superficially (on dimensional
grounds) each term in the series appears to be equally important. 
Secondly, we have not yet accounted for the effect of the exponential
derivatives. In fact we will see in the next section that when
$\ep\nu^3\sim\Cal{O}(1)$, it is these terms that impose stricter
conditions on the series truncations. For now, however, we have no
guidance other than the fact that if we account for one term of order
$\sim\ep^n\nu^n$, then we should account for \emph{all} terms at this 
order. Given this, note that for $\ep\nu^3\sim\Cal{O}(1)$ we have
$\nu^{-2}\sim\ep\nu$, and hence the terms arising from
\eqns{notes-eq12b} and \eqref{notes-eq12c} are of order
$\sim\ep^2\nu^2$. To consistently retain them, we must
therefore also retain the term linear in $\eps_2$ and the one
quadratic in $\eps_1$, when expanding the
exponential. These involve the following quantities:
\begin{subequations}
\begin{align}
\sum_{j=1}^n{(1-\frac{t_j}{t})}
\sum_{k,l,m=1}^n{\int_{-\infty}^{\delc}{d\del_1\ldots
    d\del_n\p_j\p_k\p_l\p_mW^{\rm gm}}} &= 
-\left(\frac{2}{\pi}\right)^{1/2}\frac{1}{t^2}\nu\, e^{-\nu^2/2} \,,
\la{notes-eq14a}\\
\sum_{j,k=1}^n{(1-\frac{t_j}{t})(1-\frac{t_k}{t})}
\sum_{l,l_1,l_2,l_3=1}^n{\int_{-\infty}^{\delc}{d\del_1\ldots   
    d\del_n\p_j\p_k\p_l\p_{l_1}\p_{l_2}\p_{l_3}W^{\rm gm}}} &=
-\left(\frac{2}{\pi}\right)^{1/2}\frac{4}{t^3}\nu\, e^{-\nu^2/2} \,, 
\la{notes-eq14b}
\end{align}
\end{subequations}
where we have used the result \eqref{delcderivative}, and in
\eqn{notes-eq14b} also the identity 
\be
\p_\nu^3h(\nu) = -\nu\, e^{-\nu^2/2}\,.
\la{hofnu}
\ee
We now see that the result of the path integral depends only on
$\nu$. Putting things together and computing the overall $\nu$
derivative, we find the result in \eqn{f-expderiv}.

\subsection{Saddle point calculation}
\la{app-massfuncsaddlept}
To compute the action of the exponentiated derivative operators, we
start by writing the expression in square brackets in \eqn{f-expderiv}
in terms of its Fourier transform, using the relations\footnote{We are 
  using a regulator which shifts the pole at $\lam=0$ in the last
  expression in \eqn{fourier}, to $\lam=-i\alpha$ where $\alpha$ is
  real, positive and small.}
\begin{align}
e^{-\nu^2/2} &= \lamint{e^{i\lam\nu}e^{-\lam^2/2}}\,, \nonumber\\
-\nu\, e^{-\nu^2/2} &= \lamint{(i\lam)e^{i\lam\nu}e^{-\lam^2/2}}\,,
\nonumber\\ 
\nu^2e^{-\nu^2/2} &= -\lamint{(\lam^2-1)e^{i\lam\nu}e^{-\lam^2/2}}\,,
\nonumber\\
\left(\frac{\pi}{2}\right)^{1/2}\erfc{\frac{\nu}{\sqrt{2}}} &= 
\lamint{\frac{i}{\lam}e^{i\lam\nu}e^{-\lam^2/2}} \,.
\la{fourier}
\end{align}
Together with the identity
%
$e^{A (-d/d \nu)^n} e^{i \lam \nu} = e^{A (- i \lam)^n} e^{i \lam \nu}$\,,
%
for constant $A$ and $B$, this gives
\be
f(\nu) = \left(\frac{2}{\pi}\right)^{1/2} \nu
\lamint{e^{i\lam\nu}e^{-\lam^2/2+(-i\lam)^3\eps_1/6+
    (-i\lam)^4\eps_2/24+\ldots} \Cal{P}(\lam)}
\la{FT-1}
\ee
where $\Cal{P}(\lam)$ is the truncated series given by
\be
\Cal{P}(\lam) = 1 +\frac{1}{4}i\eps_1\lam +
\frac{5}{16}\frac{i\eps_1}{\lam} -
\frac{1}{4}\lam^2\left(\frac{\eps_1^2}{2} -
\frac{\eps_2}{3}\right) + \ldots
\la{P}
\ee
The integral in eq. \eqref{FT-1} can be performed using the saddle
point approximation. We write it as
\be
f(\nu) = \left(\frac{2}{\pi}\right)^{1/2} \nu \lamint{e^{\phi(\lam)}}\,,
\la{notes-eq21}
\ee
where
\be
\phi(\lam) \equiv i \lam\nu   - \frac{1}{2}\lam^2
+ \frac{i\eps_1}{6}\lam^3 + \frac{\eps_2}{24}\lam^4 + \ln\Cal{P}(\lam)
+\ldots  
\la{notes-eq22}
\ee
The location of the saddle point, $\lam=\lam_\ast$, is
the solution of $\phi^\prime(\lam_\ast) = 0$,
and the saddle point approximation then tells us that
\be
\lamint{e^{\phi(\lam)}} = e^{\phi(\lam_\ast)}
(|\phi^{\prime\prime}(\lam_\ast)|)^{-1/2}\,,
\la{saddlepoint}
\ee
(see Appendix
\ref{append-saddlepoint} for a discussion of the errors introduced by
this approximation). 
It turns out that in
order to obtain $f(\nu)$ correctly up to order $\sim\ep^2\nu^2$, we
only need $\lam_\ast$ correct up to order $\sim\ep\nu$. The expression
for  $\phi^\prime$ at the relevant order is,
\be
\phi^\prime(\lam) = i \nu - \lam + \frac{i\eps_1}{2}\lam^2
+ \ldots,
\la{notes-eq23}
\ee
and solving for $\lam_\ast$ perturbatively up to order 
$\ep\nu$, we find 
\be
\lam_\ast = i\nu \left[1 - \frac{1}{2}\eps_1\nu 
  +\Cal{O}(\ep^2\nu^2) \right]\,.
\la{notes-eq24}
\ee
This leads to the expression for $f(\nu)$ in \eqn{f-consteps}.

\subsection{Result with full unequal time terms}
\la{app-unequaltimefull}
In the more realistic case of slowly-varying $\eps_n$, we
choose to parametrize the coefficients $\Cal{G}_3$ and $\Cal{G}_4$
(see \eqn{G3def}) in a convenient way as follows : 
\begin{align}
\Cal{G}_3^{(1,0,0)} = \frac{1}{2}\eps_1(t) c_1(t) t^{1/2}~~&;~~
\Cal{G}_3^{(2,0,0)} = -  \frac{1}{4}\eps_1(t)c_2(t) t^{-1/2} \,,
\nonumber\\
\Cal{G}_3^{(1,1,0)} =  \frac{1}{4}\eps_1(t)c_3(t) t^{-1/2} ~~&;~~
\Cal{G}_4^{(1,0,0,0)} =  \frac{1}{2}\eps_2(t)c_4(t) t\,,
\la{cndef}
\end{align}
where the coefficients $c_n(t)$ are smoothly varying functions and
depend on the NG model. They are defined in such a way
that they all reduce to unity in the toy model defined
by~\eqn{toymodel}. \fig{fig-cn} shows the behaviour of $c_1$, $c_2$
and $c_3$ with $\sig^2$, for the local and equilateral models.
The $\eps_n$ and $c_n$ are independent of redshift by construction,
since the linear growth rate $D(z)$ always drops out in their
definitions. Further, the $c_n$ do not depend on the values of
\fnl\ and \gnl. 
One can then use the definitions of $\eps_1$ and the $c_n$ to prove
the following useful relations
\begin{align}
\frac{d\ln\eps_1}{d\ln t}&=\frac{3}{2}\left(c_1-1\right)~;~ \frac{d\ln
  c_1}{d\ln t} = 1
-\frac{3}{2}c_1 + \frac{1}{c_1}\left(c_3-\frac{1}{2}c_2 \right) \,. 
\la{ep1dot-c1dot}
\end{align}
The calculation of the mass function for this general case 
proceeds completely analogously to that for the toy model, apart from
a few subtleties which we will discuss later. 
\begin{figure}[t]
\centering
\subfloat[]{\includegraphics[width=0.47\textwidth]{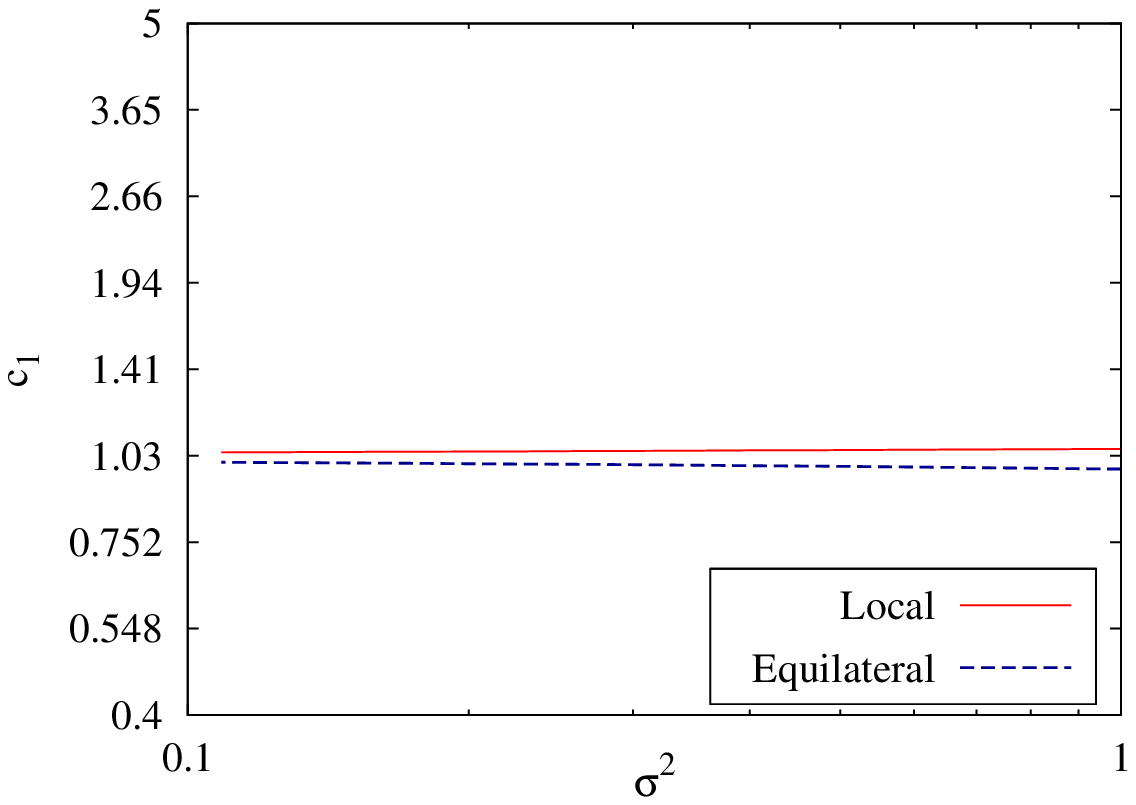}}
\la{fig-c1}\\
\subfloat[]{\includegraphics[width=0.47\textwidth]{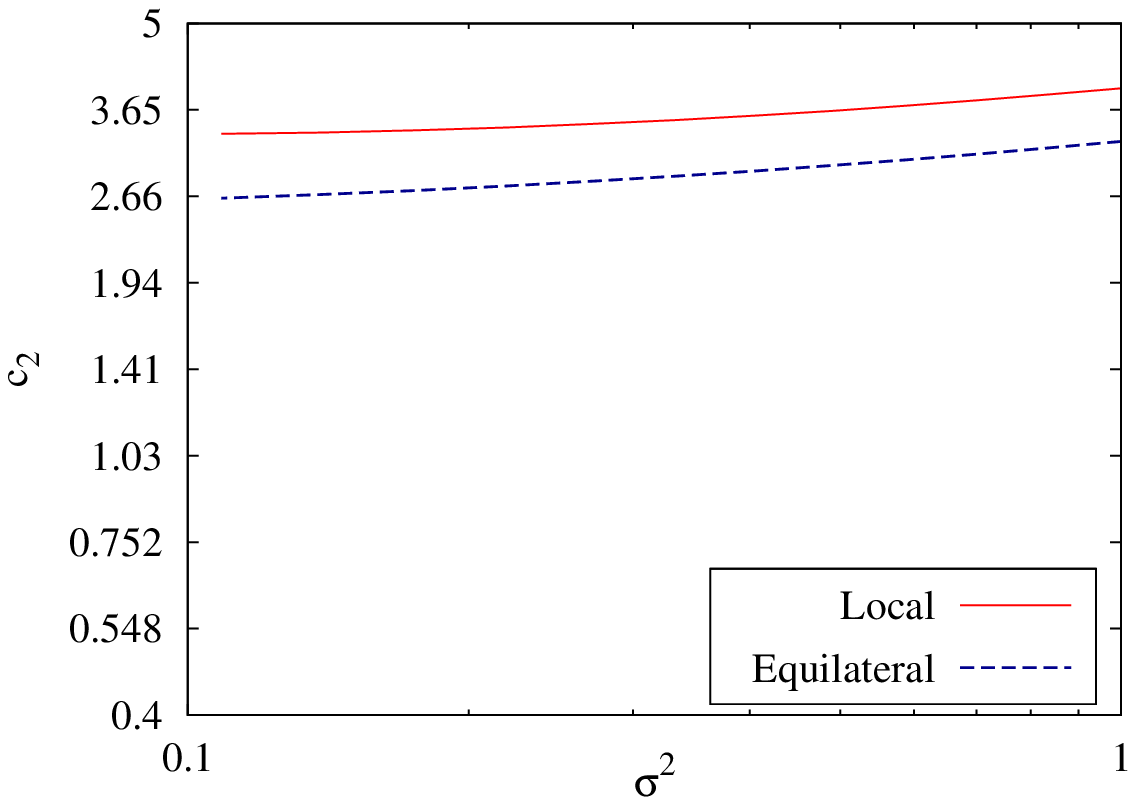}}
\la{fig-c2}
\hspace{0.03\textwidth}
\subfloat[]{\includegraphics[width=0.47\textwidth]{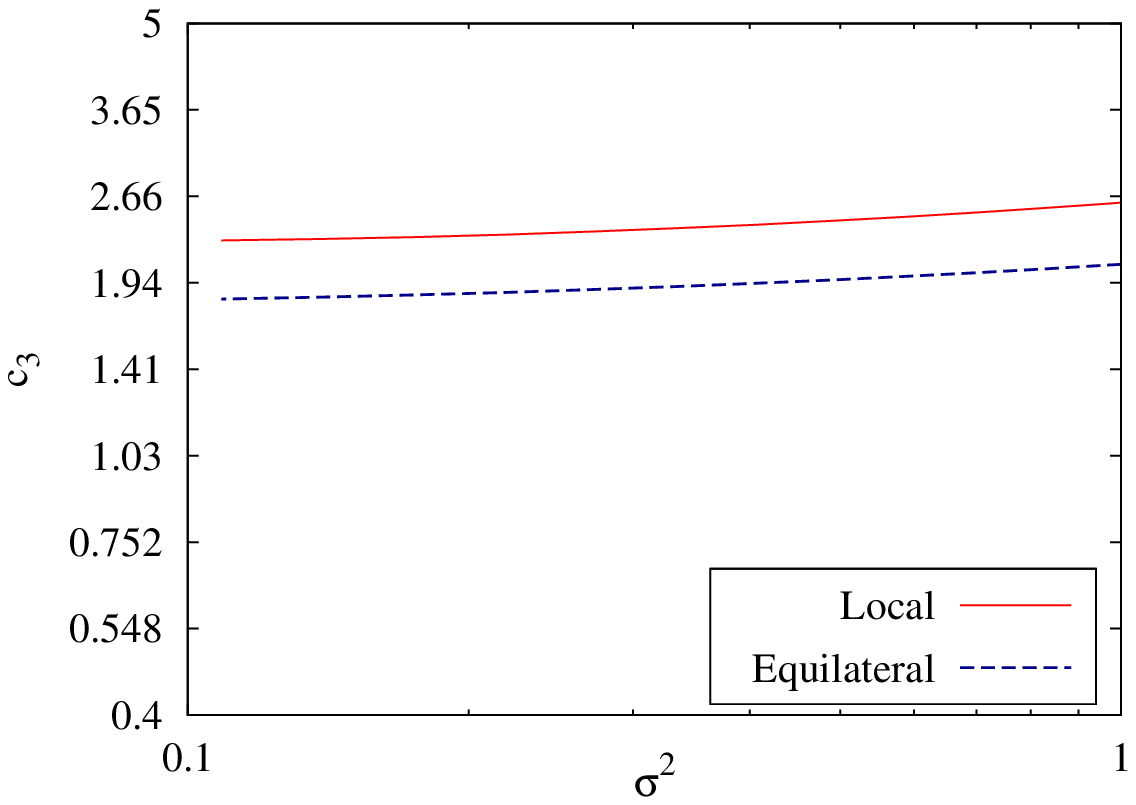}}
\la{fig-c3}
\caption{\footnotesize
The derivative coefficients $c_1$ (panel (a)), $c_2$ (panel (b)) and
$c_3$ (panel (c)), as a function of $\sig^2$, for local and equilateral
NG models. These quantities are independent of redshift and
the NG amplitudes \fnl\ and \gnl. The axes are logscale.}
\la{fig-cn}
\end{figure}
In this case \eqn{f-inter} is replaced with
\begin{align}
f &= (\nu \p_\nu - 2 t \p_t) e^{-(\eps_1(t)/3!)\p_\nu^3
  +(\eps_2(t)/4!)\p_\nu^4+\ldots} g(\nu, t) \nonumber \\
&= \left[ \nu + \frac{1}{3} \frac{d\ln\eps_1}{d\ln t} \eps_1 \p_\nu^2 
  - \frac{1}{12} \frac{d\ln\eps_2}{d\ln t} \eps_2 \p_\nu^3 \right]
e^{-(\eps_1(t)/3!)\p_\nu^3 +(\eps_2(t)/4!)\p_\nu^4+\ldots}\, \p_\nu
g(\nu, t) \nonumber\\
&\ph{\left[ \nu + \frac{1}{3} \frac{d\ln\eps_1}{d\ln t} \eps_1
    \p_\nu^2 - \frac{1}{12} \frac{d\ln \eps_2}{d\ln t} \eps_2 \p_\nu^3
    \right]} - 2t e^{-(\eps_1(t)/3!)\p_\nu^3+(\eps_2(t)/4!)\p_\nu^4+\ldots}
\,\p_t g(\nu,t)\,,
\la{f-inter-new}
\end{align}
%
wher the function $g(\nu,t)$ can be shown to be  
\begin{align}
g(\nu, t) &= \left( \frac{2}{\pi} \right)^{1/2}
\bigg[ \left( \frac{\pi}{2} \right)^{1/2}  \erf{\frac{\nu}{\sqrt{2}}} 
  + \frac{1}{4}\eps_1 c_1 e^{-\nu^2/2}
  + \frac{\eps_1}{4} \left(\frac{3}{4} c_2 - 2 c_3 \right) h(\nu)
  \nonumber\\
&\ph{\left(\frac{2}{\pi}\right)^{1/2} \nu e^{-(\eps_1/3!)\p_\nu^3 +
  (\eps_2/4!)\p_\nu^4 +\ldots}\p_\nu\bigg[\left(\frac{\pi}{2}\right)^{1/2}} 
    -\frac{1}{8} \eps_1^2 c_1^2 \nu e^{-\nu^2/2} + \frac{1}{12} \eps_2
    c_4 \nu e^{-\nu^2/2} +\ldots \bigg] \, ,
\la{g}
\end{align}
The expression in \eqn{f-inter-new} can be evaluated analogously to
\eqn{f-expderiv}, since the additional derivatives pose no conceptual
difficulty. The result of the saddle point calculation, correct up to
quadratic order assuming $\ep\nu^3\sim\Cal{O}(1)$, and after using the 
relations \eqref{ep1dot-c1dot}, is given in \eqn{f-geneps}.

\section{Truncation of the perturbative series}
\la{app-truncation}
\subsection{Analysis of ``transition points''}
\la{app-transitionpts}
In this appendix we analyse the consistency of our series truncation
at various mass scales. When $\ep\nu^3\simeq1$, in the  polynomial in
\eqref{mf-symbolic-start} we retain the terms $\ep\nu\simeq\nu^{-2}$,
$(\ep\nu^{-1},\ep^2\nu^2)\simeq\nu^{-4}$, and we discard
$(\ep\nu^{-3},\ep^2,\ep^3\nu^3)\simeq\nu^{-6}$. It would seem that our
expression is then correct upto order $\sim\nu^{-4}$. However, the
terms discarded in the exponential have the form
$\exp(\Cal{O}(\ep^3\nu^5))\sim\exp(\Cal{O}(\nu^{-4}))\sim1 +  
\Cal{O}(\nu^{-4})$. The error we are making is thus of the same
order as the smallest terms we are retaining, and it therefore makes
sense to \emph{also} ignore all the terms of order $\sim\nu^{-4}$ which
we computed in the polynomial. The consistent expression when
$\ep\nu^3\simeq1$ is then given by
\be
f\sim e^{- \frac12 \nu^2 \left(1 + \ep\nu + \ep^2\nu^2\right)} \left[1+ \ep\nu
  + \Cal{O}\left(\nu^{-4}\right) \right]\,.
\la{mf-symbolic-epnu3-0}
\ee
Clearly, similar arguments can be applied at smaller scales where,
e.g. one might have $\ep\nu^3\simeq\nu^{-1},\nu^{-2}$, etc. It is then 
important to ask which mass scales correspond to these ``transition
points''. In \fig{nuvsM} we plot $\nu(M,z)$ given by \eqn{nuofz} in an
observationally interesting mass range, for three different
redshifts. The horizontal lines mark the 
transition points where $\ep\nu^3$ becomes equal to (from top to
bottom) $1$, $\nu^{-1}$, $\nu^{-2}$, $\nu^{-3}$, $\nu^{-4}$
and $\nu^{-5}$. We fix $\ep=1/300$ which follows from the fact that in
the local model with $\fnl=100$ we have $\eps_1\simeq0.02$ (see
\fig{fig-eps}), and the expression for $f(\nu,M)$ contains the
quantity $\eps_1/6$ in the exponential. From the intersections of the
horizontal lines with the curves, we see that different
transition points are relevant at different redshifts, and their
locations also obviously depend on the value of $\ep$. For example, we
find that the transition point where $\ep\nu^3\simeq\nu^{-2}$, remains
accessible even when $\ep$ is an order of magnitude smaller (with
$\ep\simeq1/3000$, this transition occurs at $\nu\simeq4.96$). The
transitions at $\ep\nu^3\simeq1,\nu^{-1}$ on the other hand, are not
accessible for this level of NG. The transition at
$\ep\nu^3\simeq\nu^{-2}$ is therefore observationally very
interesting. 
\begin{figure}[t]
\centering
\includegraphics[height=0.25\textheight]{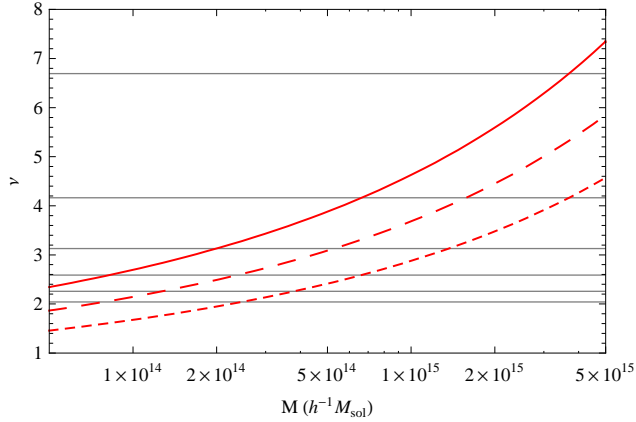}
\caption{\footnotesize $\nu\equiv\delc(z)/\sig(M)$ in the range
  $5\cdot10^{13}<(M/h^{-1}\Msol)<5\cdot10^{15}$ for
  three different redshifts, with $\ep=1/300$. The solid, long
  dashed and short dashed curves correspond to redshifts $z=1$, $0.5$
  and $0$ respectively. The horizontal lines mark the transition
  points where $\ep\nu^3$ becomes equal to (from top to bottom) $1$,
  $\nu^{-1}$, $\nu^{-2}$, $\nu^{-3}$, $\nu^{-4}$ and $\nu^{-5}$.} 
\la{nuvsM}
\end{figure}

We will now discuss in some detail the truncation of our expression
for $f$, at various transition points. The goal is to try and settle on a
\emph{single} expression which is valid over a wide range of scales
(i.e. across several transition points). This can then be applied
without worrying about truncation inconsistencies. Of course, the
order of the discarded terms will then depend on the particular
transition point being considered, leading to a scale dependent
theoretical error. 
\subsubsection{$\ep\nu^3\simeq\nu^{-1}$}
\noindent
At this transition point, the terms we retain in the exponential are 
\be
\ep\nu^3\simeq\nu^{-1} ~~;~~ \ep^2\nu^4\simeq\nu^{-4}\,, 
\nonumber
\ee
while discarding
$\Cal{O}(\ep^3\nu^5)=\Cal{O}(\nu^{-7})$. In the polynomial meanwhile,
we retain 
\be
\ep\nu\simeq\nu^{-3}~~;~~ \ep\nu^{-1}\simeq\nu^{-5} ~~;~~
\ep^2\nu^2\simeq\nu^{-6}\,, 
\nonumber
\ee
while discarding
\be
\Cal{O}(\ep\nu^{-3})= \Cal{O}(\nu^{-7}) ~~;~~ \Cal{O}(\ep^2)=
\Cal{O}(\nu^{-8})~~;~~ \Cal{O}(\ep^3\nu^3)= \Cal{O}(\nu^{-9})\,.
\nonumber
\ee
Our expression \eqref{mf-symbolic-start} therefore retains all terms
correctly up to order $\sim\nu^{-6}$, and is consistent. With some
foresight however, it turns out to be more convenient to degrade this
expression somewhat by also discarding the polynomial quadratic term
$\ep^2\nu^2\simeq\nu^{-6}$. The remaining expression,
\be
f\sim e^{- \frac12 \nu^2 \left(1 + \ep\nu + \ep^2\nu^2 \right)}  \left[1+
  \ep\nu + \frac{\ep}{\nu} 
  + \Cal{O}\left(\nu^{-6}\right) \right]\,,
\la{mf-symbolic-epnu3-1}
\ee
is also consistent at this transition point, and has a form which is
identical to the ones we will see next.

\subsubsection{$\ep\nu^3\simeq\nu^{-2}$}
\noindent
As we mentioned earlier, this transition point is observationally
quite interesting. The terms we retain in the exponential are 
\be
\ep\nu^3\simeq\nu^{-2} ~~;~~ \ep^2\nu^4\simeq\nu^{-6}\,, 
\nonumber
\ee
while discarding
$\Cal{O}(\ep^3\nu^5)=\Cal{O}(\nu^{-10})$, and in the polynomial we
retain
\be
\ep\nu\simeq\nu^{-4}~~;~~ \ep\nu^{-1}\simeq\nu^{-6} ~~;~~
\ep^2\nu^2\simeq\nu^{-8}\,, 
\nonumber
\ee
while discarding
\be
\Cal{O}(\ep\nu^{-3})= \Cal{O}(\nu^{-8}) ~~;~~ \Cal{O}(\ep^2)=
\Cal{O}(\nu^{-10})~~;~~ \Cal{O}(\ep^3\nu^3)= \Cal{O}(\nu^{-12})\,.
\nonumber
\ee
This time we see that the term $\ep\nu^{-3}$ has become as important
as the quadratic term $\ep^2\nu^2$ in the polynomial, and to be
consistent we should discard the quadratic term. The expansion should
read 
\be
f\sim e^{- \frac12 \nu^2 \left(1 + \ep\nu + \ep^2\nu^2\right)} \left[1+ \ep\nu
  + \frac{\ep}{\nu} + \Cal{O}\left(\nu^{-8}\right) \right]\,.
\la{mf-symbolic-epnu3-2}
\ee

\subsubsection{$\ep\nu^3\simeq\nu^{-3}$}
A similar analysis as above shows that at this stage
$\ep\nu^{-3}\simeq\nu^{-9} > \ep^2\nu^2$, and a consistent expression
again requires dropping the quadratic term in the polynomial, leaving 
\be
f\sim e^{- \frac12 \nu^2 \left(1 + \ep\nu + \ep^2\nu^2\right)}  \left[1+
  \ep\nu + \frac{\ep}{\nu} + \Cal{O}\left(\nu^{-9}\right) \right]\,.
\la{mf-symbolic-epnu3-3}
\ee
\subsubsection{$\ep\nu^3\simeq\nu^{-4}$ and smaller}
Beyond this point, the term $\ep\nu^{-3}$ which we discard in the
polynomial, becomes comparable or larger than the quadratic term of
the exponential as well, and a consistent expression becomes
\be
f\sim e^{- \frac12 \nu^2 \left(1 + \ep\nu \right)}  \left[1+ \ep\nu +
  \frac{\ep}{\nu} + \ldots \right]
\la{mf-symbolic-epnu3-gtr3}
\ee
The parametric order of the terms now discarded, depends on the exact
relation between $\ep\nu^3$ and $\nu^{-1}$.
\vskip 0.1in
Finally, note that the error introduced by setting $\nu\to\nu_{\rm g}$
where $\nu_{\rm g}$ is defined using the variance of a \emph{Gaussian}
field, was estimated in section \ref{calc-f} as $\Cal{O}(\ep^2)$. When
$\ep\nu^3\simeq1$, this error is of order $\Cal{O}(\ep^3\nu^3)$ and can
therefore be consistently ignored. It is not hard to see that at
\emph{all} lower transition points, this error continues to be
comparable to or smaller than the largest terms being discarded, and
can hence be consistently ignored. This finally leads to the
conclusion stated in the main text.

\subsection{Truncation in MR and LMSV results}
\la{app-truncateothers}
Since the mass scale where $\ep\nu^3\simeq1$ is on the border of the
observed mass window (for galaxy cluster observations), even at high
redshifts, let us therefore directly look at the case
$\ep\nu^3\simeq\nu^{-2}$ which, as we saw, is accessible over a wide
range of redshifts for $\ep\sim10^{-2}$, and at high redshifts also for
$\ep\sim10^{-3}$. In this case the terms MR and LMSV retain have
magnitudes  
\be
\ep\nu^3\simeq\nu^{-2}~;~~ \ep\nu\simeq\nu^{-4} ~;~~
\ep\nu^{-1}\simeq\nu^{-6}\,,
\nonumber
\ee
and terms like $\ep\nu^{-3}\simeq\nu^{-8}$ are discarded. We know from
our expression however, that $\ep\nu^3$ appears in the exponential,
and therefore leads to terms like $(\ep\nu^3)^2\simeq\nu^{-4}$ and
$(\ep\nu^3)^3\simeq\nu^{-6}$ when the exponential is expanded, which
are of the same order as the terms retained in
\eqref{mf-symbolic-MR}. The exponential also contributes a term
$\ep^2\nu^4\simeq\nu^{-6}$, which in fact involves the
\emph{tri}spectrum of NG, again at the order retained
by MR and LMSV. The error in the expression \eqref{mf-symbolic-MR} when
$\ep\nu^3\simeq\nu^{-2}$, is therefore $\Cal{O}(\ep\nu)$. (A similar
analysis shows that the error at transition point where
$\ep\nu^3\simeq\nu^{-1}$, is $\Cal{O}(\nu^{-2})>\Cal{O}(\ep\nu)$.)

From a purely parametric point of view, the situation for MR and LMSV
improves as $\nu$ is decreased further, and the expression
\eqref{mf-symbolic-MR} as it stands, becomes exactly consistent (in
the sense discussed in the previous subsection, see below
\eqn{mf-symbolic-final}) when $\ep\nu^3\simeq\nu^{-5}$, because at
this stage $\ep\nu^{-1}\simeq\nu^{-9}$ while
$(\ep\nu^3)^2\simeq\nu^{-10}$  and $\ep^2\nu^4\simeq\nu^{-12}$, and
hence the exponential only 
contributes a single linear term $\ep\nu^3$. More importantly, LMSV's
expression also has errors due to the absence of the unequal time
terms discussed earlier, which are of order $\sim\ep\nu$ and can be
dominant over the others. For the intermediate
transitions, the analysis shows that when $\ep\nu^3\simeq\nu^{-3}$,
the error in \eqref{mf-symbolic-MR} is
$\Cal{O}(\nu^{-6})>\Cal{O}(\ep\nu^{-1})$, and when
$\ep\nu^3\simeq\nu^{-4}$, the error is $\Cal{O}(\ep\nu^{-1})$. This
should be compared with our result \eqref{f-final}, in which the error
(at least on large scales) is always parametrically \emph{smaller}
than the smallest terms we retain.

\section{Hierarchy of terms in \eqn{f-inter}}
\la{append-incompgamma}        
Here we argue why the hierarchy of terms ordered by powers of
$\nu^{-2}$ emerges on expanding the exponentiated derivative operators
in \eqn{f-inter}. Focusing on terms involving the $3$-point
correlator, one sees that a generic term in the expansion contains
some powers of $(\eps_1t^{3/2})$, multiplying an $n$-dimensional
integral containing some summations
$\sim\sum_{j_1,j_2,\ldots=1}^n(1-t_{j_1}/t)^{p_1}(1-t_{j_2}/t)^{p_2}
\ldots \p_{j_1}\p_{j_2} \ldots$, and also some summations over
``free'' derivatives $\sim \sum_{k_1,k_2 \ldots
  =1}^n\p_{k_1}\p_{k_2}\ldots$, all of this acting on $W^{\rm
  gm}$. More precisely, the structure of the terms is
\begin{align}
&\sim (\eps_1t^{3/2})^m \sum_{j_1,..,j_{3m}}\int_{-\infty}^{\delc}
d\del_1\ldots d\del_n
\left[\left(1-t_{j_1}/t\right) \ldots 
  \left(1-t_{j_m}/t\right) \right]^p
\left[\left(1-t_{j_{m+1}}/t\right) \ldots 
  \left(1-t_{j_{2m}}/t\right) \right]^q
\nonumber\\
&\ph{\sim (\eps_1t^{3/2})^m
  \sum_{j_1,..,j_{3m}}\int_{-\infty}^{\delc} d\del_1\ldots d\del_n} 
\times
\left[\left(1-t_{j_{2m+1}}/t\right) \ldots 
  \left(1-t_{j_{3m}}/t\right) \right]^r
\p_{j_1}\ldots\p_{j_{3m}} \, W^{\rm gm}\,,
\la{type123terms}
\end{align}
for $m\geq1$ and non-negative $p,q,r$ such that not all three are
zero. The terms we have considered in the text are $(m,p,q,r) =
(1,1,0,0)$, $(1,1,1,0)$, $(1,2,0,0)$ and $(2,1,0,0)$. We have already
discussed how the ``free'' derivatives can be pulled out of the
integral and converted to $\p_\nu$. For the ``non-free'' derivatives,
we see that what is important is the \emph{total} number of
$(1-t_j/t)$ factors accompanying these derivatives. For example, the
$(1,1,1,0)$ term in \eqn{notes-eq12c} has the same structure as the
$(1,2,0,0)$ term in \eqn{notes-eq12b} -- the effect of
$\sum_{j,k}(1-t_j/t)(1-t_k/t)\p_j\p_k$, up to numerical factors, is
identical to that of $\sum_{j,k}(1-t_j/t)^2\p_j\p_k$. This is expected
to be true also with higher numbers of non-free
derivatives. 

It is then possible to understand the hierarchy of terms by only
considering terms containing $\sum_j(1-t_j/t)^p\p_j$, and no other
non-free derivatives. The basic object to study now becomes
\be
\sum_j(1-t_j/t)\int d\del_1\ldots d\del_n\p_jW^{\rm gm}\,, 
\nonumber
\ee
which in the continuum limit can be shown to reduce to the integral
\be
g_{(0)}\left(\frac{\nu^2}{2}\right) \equiv
\int_0^1\frac{dy}{y^{3/2}}(1-y)^{1/2}e^{-\nu^2/2y} =
\frac{\sqrt{\pi}}{2}\Gam\left(-\frac{1}{2},\frac{\nu^2}{2}\right) \,.
\la{heir-0}
\ee
Notice the similarity with the function $h(\nu)$ in \eqn{hofnu}, which
of course is not accidental given the definitions of these
objects. It is now easy to check that increasing the powers of
$(1-t_j/t)$ in the summation amounts to increasing the powers of
$(1-y)$ in $g_{(0)}$. We are then comparing (with $A=\nu^2/2$)
$g_{(0)}(A)$ with $g_{(p)}(A)$ where
\be
g_{(p)}(A) \equiv \int_0^1\frac{dy}{y^{3/2}}(1-y)^{1/2+p}e^{-A/y}\,.
\la{heir-p}
\ee
Starting with $p=1$ and manipulating the integrals, it is
straightforward to establish the recurrence
\be
g_{(p+1)}(A) = g_{(p)}(A) - \int_A^\infty d\tilde A\,g_{(p)}(\tilde
A)\,. 
\la{heir-recur}
\ee
The argument is now almost complete. We know that for large
$A=\nu^2/2$, we have
$\Gam(n,A)=e^{-A}A^{n-1}(1+\Cal{O}(A^{-1}))$. Hence
$g_{(0)}(A)=(\sqrt{\pi}/2)A^{-3/2}e^{-A} (1+ \Cal{O}(A^{-1}))$, and
its integral from $A$ to $\infty$ gives a leading term proportional to
$\Gam(-3/2,A)=e^{-A}A^{-5/2}(1+\Cal{O}(A^{-1}))$. The pattern is now
clear: $g_{(p)}(A) \sim A^{-3/2-p}e^{-A}(1+\Cal{O}(A^{-1}))$, and
since $A=\nu^2/2$, this explains the hierarchy of terms in powers of
$\nu^{-2}$, in \eqn{f-inter}. 

\section{The saddle point approximation}
\label{append-saddlepoint}
\noindent
In this appendix we discuss the saddle point approximation of the
integrals of the type appearing in section~\ref{gaussmark-intro}, and
estimate the error it induces. We will argue that the errors
introduced by the saddle point approximation are much smaller than
those due to truncating the perturbative series in the small parameters
$\ep$ and $\nu^{-1}$.
For an introduction to the saddle point approximation see~\Cite{Erdelyi}.
Since we only wish to discuss the saddle point method
in this appendix, we will ignore here the complications introduced by
the unequal time correlators, i.e. in \eqn{FT-1} we set $\Cal{P}(\lam)=1$.
We will also work here to first order in $\ep\nu$.
The extension to a more general case is straightforward
and the result is given by \eqref{f-geneps} as
described in section \ref{gaussmark-intro}. We begin with expression
\eqref{FT-1}:
\begin{equation}
f(\nu) =
\bigg(\frac{2}{\pi}\bigg)^{1/2}\nu\int^\infty_{-\infty}
\frac{\mathrm{d}\lambda}{\sqrt{2\pi}}e^{g(\lambda)}\,,  
\label{saddlepoint:integral}
\end{equation}
where $g(\lambda)\equiv i\nu\lambda-\lambda^2/2 +
(-i\lambda)^3\eps_1/6 + \mathcal{O}(\ep^2\lambda^4)$. 

We first find the location of a saddle point $\lambda_\ast$ of the
function $g(\lambda)$, by perturbatively solving $g'(\lambda_\ast) =
0$ using $\ep\nu$ as the small parameter and demanding
$g''(\lambda_\ast)<0$. The first-order solution is
\begin{equation}
\lambda_\ast = i\nu\big(1
- \eps_1\nu/2 + \mathcal{O}(\ep^2\nu^2)\big)\;,
\end{equation}
\begin{equation}
g(\lambda_\ast) = -\frac{\nu^2}{2}(1-\frac{1}{3}\eps_1\nu +
\mathcal{O}(\ep^2\nu^2))\;, 
\end{equation}
\begin{equation}
g''(\lambda_\ast) = -1 -\eps_1\nu +
\mathcal{O}(\ep^2\nu^2)\;.
\label{saddlepoint:g''}
\end{equation}
The saddle point approximation consists roughly of performing a Taylor
expansion of $g(\lambda)$ to second order around $\lambda_\ast$ in the
integrand of~\eqref{saddlepoint:integral} and performing the resulting
Gaussian integral. We will carry this out explicitly below.
The saddle point prescription will give a good approximation to the integral as
long as $g(\lambda)$ attains a global maximum at $\lambda_\ast$ (along
the contour of integration); this is indeed our case since 
the integrand in \eqn{saddlepoint:integral} will be nearly a Gaussian
centered at $\lambda_\ast$ in the complex plane.

Notice that $\mathrm{Im}\;\lambda_\ast\neq 0$, requiring a deformation
of the contour of integration such that it passes through
$\lambda_\ast$. The deformation of the path of integration can be
performed by taking a closed contour formed by four pieces: The real
axis $C_1$, the line $\mathrm{Im}\;\lambda = \mathrm{Im}\;
\lambda_\ast$ which we call here $-C_2$, and the closures of this
contour at possitive and negative infinity. The integral in this
closed contour must be zero, and since the integral on the closures of
the contour at infinity can be assumed to vanish, we have $\int_{C_1}
= \int_{C_2}$. Therefore $C_2$ is the desired deformation of the
contour which passes through $\lambda_\ast$\footnote{Technically, one
  should also require that $\mathrm{Im}\;g(\lambda)$ be nearly
  constant along the deformed contour for the saddle point
  approximation to work. In our case one can show that
  $\mathrm{Im}\;g$ will be suppressed by $\epsilon$.  This and all
  errrors induced by the saddle point are accounted for in equation
  \eqref{saddlepoint:interror}.
}.  We can then make a series of approximations in the integral
\eqref{saddlepoint:integral}, which we discuss below,
\begin{align}
  \lamint{e^{g(\lambda)}} &\approx
  \int_{-\infty}^{\infty}
  \frac{d\lambda}{\sqrt{2\pi}} \; 
  e^{g(\lambda_\ast)+g''(\lambda_\ast)(\lambda-\lambda_\ast)^2/2}
  \nonumber \\
  &=
  e^{g(\lambda_\ast)}\left(-g''(\lambda_\ast)\right)^{-1/2}\nonumber
  \\
  &= e^{-\frac12 \nu^2 \left(1-\eps_1\nu/3 +\mathcal{O}(\ep^2\nu^2)\right)}
  \left(1 + \eps_1\nu + \mathcal{O}(\ep^2 \nu^2)\right)^{-1/2}\;.
\label{saddlepoint:intapprox}
\end{align}
Here the integrations are performed along the deformed contour.

In order to estimate the errors induced by the approximation done in
equation \eqref{saddlepoint:intapprox}, one can keep higher orders in
the Taylor expansion of the function in the exponential:
\begin{align}
  \lamint{e^{g(\lambda)}} &\approx
  \int_{-\infty}^{\infty}
  \frac{d\lambda}{\sqrt{2\pi}} \; 
  e^{g(\lambda_\ast)+g''(\lambda_\ast)(\lambda-\lambda_\ast)^2/2 +
  g^{(3)}(\lambda_\ast)(\lambda - \lambda_\ast)^3/6 +
  g^{(4)}(\lambda_\ast)(\lambda - \lambda_\ast)^4/24 + \dots}
  \nonumber \\
  &\approx
  e^{g(\lambda_\ast)}\left(-g''(\lambda_\ast)\right)^{-1/2}
  \nonumber \\
  &+\: \int_{-\infty}^{\infty} \frac{\mathrm{d}z}{\sqrt{2\pi}}
  \bigg\{\frac{1}{6}g^{(3)}(\lambda_\ast)z^3 +
  \frac{1}{72}\big[g^{(3)}(\lambda_\ast)\big]^2z^6 +
  \frac{1}{24}g^{(4)}(\lambda_\ast)z^4 + \dots \bigg\}
  e^{g(\lambda_\ast)+g''(\lambda_\ast)z^2/2}
  \nonumber  \\
  &=   e^{g(\lambda_\ast)}\left(-g''(\lambda_\ast)\right)^{-1/2}
(1 + \mathcal{O}(\epsilon^2))\;.
\label{saddlepoint:interror}
\end{align}
Here we used the fact that $g^{(3)}(\lambda_\ast) =
\mathcal{O}(\epsilon)$ and $g^{(4)}(\lambda_\ast) =
\mathcal{O}(\epsilon^2)$. The integrals in the second equality of this
derivation can be computed analytically, which allows one to go to
arbitrary accuracy with the saddle point technique. Notice
that the results of these integrations are of higher order than the
terms we retain. In the main text, 
where the integral contains also a polynomial $\cal{P}(\lambda)$ one
can again compute the errors via similar Taylor expansions. These
errors can be shown to be of order $\mathcal{O}(\epsilon^2)$,
comparable to other terms which we ignore.

\begin{figure}[t]
\centering
\subfloat[]{
  \includegraphics[width=0.47\textwidth]{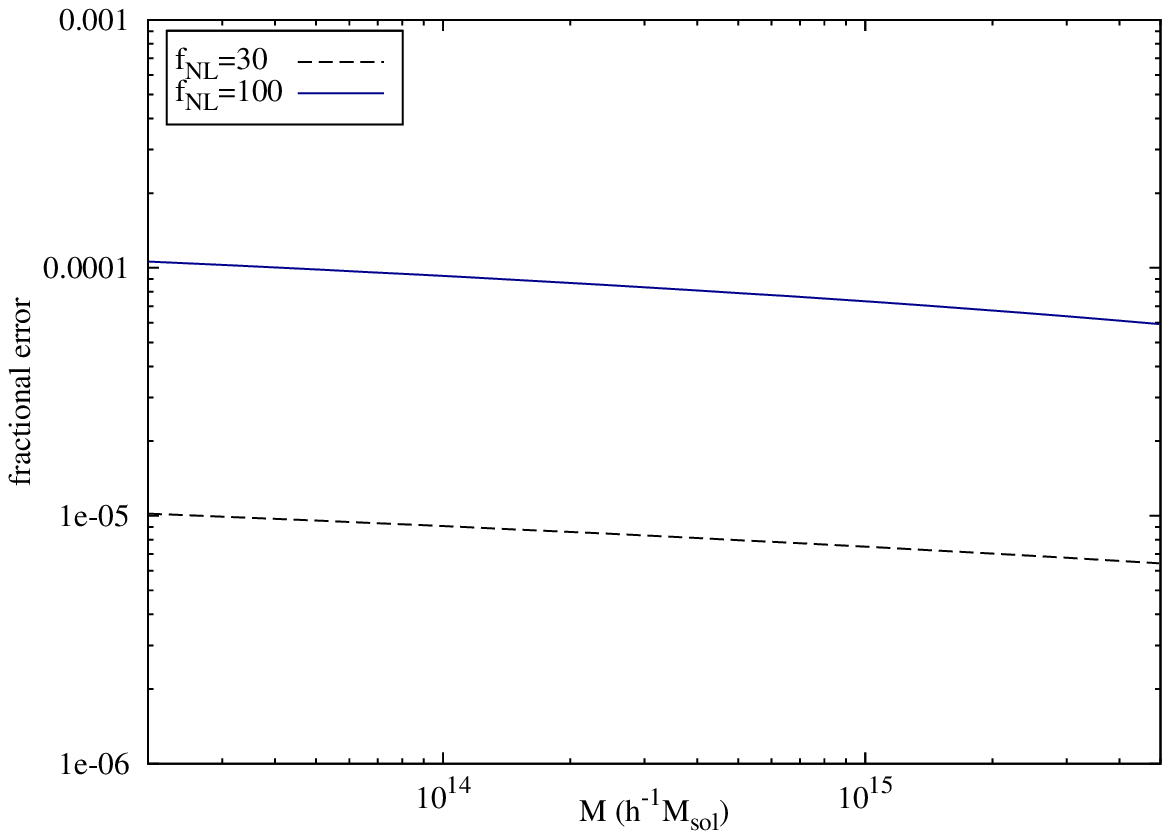}} 
\subfloat[]{
  \includegraphics[width=0.47\textwidth]{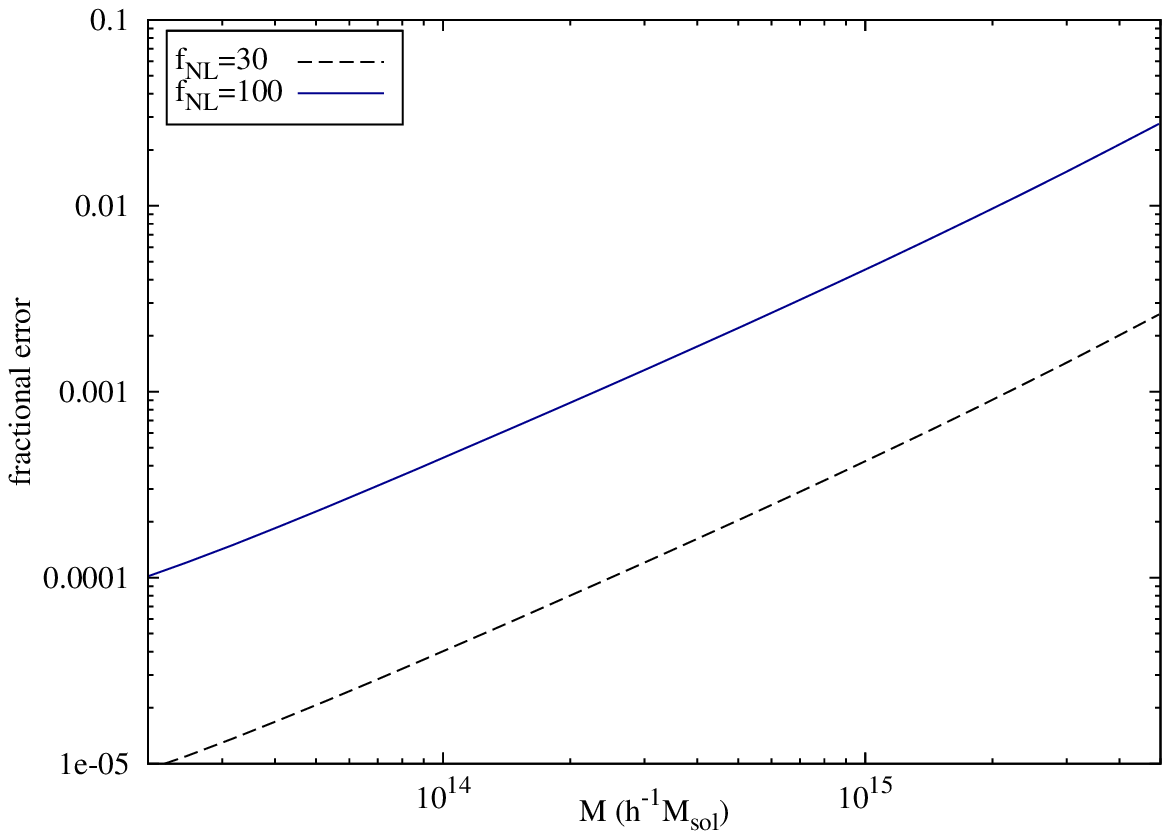}} 
\caption{\footnotesize Panel (a) : Fractional difference between the
  saddle point approximation on the r.h.s. of \eqn{saddlepoint:exact}
  and the numerical integration of the l.h.s. of the same equation. 
  Panel (b) : Total error induced on the result of the toy model
  \eqref{saddlepoint:exact} by both the saddle point approximation and
  the perturbative expansion to leading order in $\ep\nu$. We
  plot the fractional difference between the numerical integration of
  the l.h.s. of \eqn{saddlepoint:exact} and the approximation
  \eqref{saddlepoint:intapprox}. Both panels show the results for a local
  NG with two values of \fnl.}
\label{saddlepoint:toy}
\end{figure}

One can also estimate the errors introduced by our opproximations by using
the following toy model in which everything is computable:
Take the $3$-point cumulant $\eps_1$ to be different from zero and all
higher order cumulants $\eps_n$ for $n\geq2$ to be zero\footnote{This
  toy model is inconsistent because if the third cumulant is different
  from zero, then all higher cumulants must also be different from
  zero. We use it here only to estimate how good the saddle point
  prescription is in approximating an integral, and compare it with
  errors induced by a perturbative expansion in $\ep\nu$.}. For such a
model the integral is
\begin{equation}
\int_{-\infty}^\infty\mathrm{d}\lambda \;e^{i\nu\lambda-\lambda^2/2 +
(-i\lambda)^3\varepsilon_1/6} \approx
\Bigg(\frac{2\pi}{\sqrt{1+2\varepsilon_1\nu}}\Bigg)^{1/2}
\exp\Bigg(\frac{1-\sqrt{1+2\varepsilon_1\nu} 
  + \varepsilon_1 \nu
  \big(3-2\sqrt{1+2\varepsilon_1\nu}\big)}{3\varepsilon_1^2}\Bigg)\;.
\label{saddlepoint:exact}
\end{equation}
In the r.h.s of this equation we have used the saddle point
approximation but have made no expansion in $\ep\nu$. By comparing the
numerical integration of the l.h.s. with the expression on the r.h.s.
(panel (a) of \fig{saddlepoint:toy}), one can see that the errors
introduced by the saddle point approximation are indeed of order
$\epsilon^2$ as indicated by \eqref{saddlepoint:interror}. On the
other hand, one can use the numerical integration of the left hand
side of this equation and compare it with the approximation
\eqref{saddlepoint:intapprox} (panel (b) of \fig{saddlepoint:toy}), to
see that the biggest error is of order $\ep^2\nu^2$ induced by the
fact that we perform a perturbative expansion in $\ep\nu$.
Notice that here we considered only the leading order in $\ep\nu$ and
ignored unequal time correlators, while in the main text we present a
result which is more precise (to next to leading order in $\ep\nu$)
and complete (using the excursion set formalism rigorously).


\begin{thebibliography}{99}

\bibitem{Dalal:2007cu}
  N.~Dalal, O.~Dore, D.~Huterer and A.~Shirokov,
  ``The imprints of primordial non-Gaussianities on large-scale
  structure: scale dependent bias and abundance of virialized
  objects,'' 
  Phys.\ Rev.\  D {\bf 77} (2008) 123514
  [arXiv:0710.4560 [astro-ph]].

\bibitem{Matarrese:2008nc}
  S.~Matarrese and L.~Verde,
  ``The effect of primordial non-Gaussianity on halo bias,''
  Astrophys.\ J.\  {\bf 677} (2008) L77
  [arXiv:0801.4826 [astro-ph]].

\bibitem{Slosar:2008hx}
  A.~Slosar, C.~Hirata, U.~Seljak, S.~Ho and N.~Padmanabhan,
  ``Constraints on local primordial non-Gaussianity from large scale
  structure,''
  JCAP {\bf 0808} (2008) 031
  [arXiv:0805.3580 [astro-ph]].

\bibitem{Komatsu:2010fb}
  E.~Komatsu {\it et al.},
  ``Seven-Year Wilkinson Microwave Anisotropy Probe (WMAP) Observations: Cosmological Interpretation,''
  arXiv:1001.4538 [astro-ph.CO].

\bibitem{Sartoris:2010cr}
  B.~Sartoris, S.~Borgani, C.~Fedeli, S.~Matarrese, L.~Moscardini, P.~Rosati and J.~Weller,
  ``The potential of X-ray cluster surveys to constrain primordial non-Gaussianity,''
  arXiv:1003.0841 [astro-ph.CO].

\bibitem{Carbone:2008iz}
  C.~Carbone, L.~Verde and S.~Matarrese,
  ``Non-Gaussian halo bias and future galaxy surveys,''
  Astrophys.\ J.\  {\bf 684}, L1 (2008)
  [arXiv:0806.1950 [astro-ph]].

\bibitem{Carbone:2010sb}
  C.~Carbone, O.~Mena and L.~Verde,
  ``Cosmological Parameters Degeneracies and Non-Gaussian Halo Bias,''
  JCAP {\bf 1007}, 020 (2010)
  [arXiv:1003.0456 [astro-ph.CO]].

\bibitem{Cunha:2010zz}
  C.~Cunha, D.~Huterer and O.~Dore,
  ``Primordial non-Gaussianity from the covariance of galaxy cluster counts,''
  arXiv:1003.2416 [astro-ph.CO].

\bibitem{Sefusatti:2009qh}
  E.~Sefusatti,
  ``1-loop Perturbative Corrections to the Matter and Galaxy Bispectrum with
  non-Gaussian Initial Conditions,''
  Phys.\ Rev.\  D {\bf 80} (2009) 123002
  [arXiv:0905.0717 [astro-ph.CO]].

\bibitem{Jimenez:2009us}
  R.~Jimenez and L.~Verde,
  ``Implications for Primordial Non-Gaussianity ($f_{NL}$) from weak lensing masses
  of high-z galaxy clusters,''
  Phys.\ Rev.\  D {\bf 80}, 127302 (2009)
  [arXiv:0909.0403 [astro-ph.CO]].

\bibitem{Verde:2010wp}
  L.~Verde,
  ``Non-Gaussianity from Large-Scale Structure Surveys,''
  arXiv:1001.5217 [astro-ph.CO].

\bibitem{Desjacques:2010jw}
  V.~Desjacques and U.~Seljak,
  ``Primordial non-Gaussianity from the large scale structure,''
  arXiv:1003.5020 [astro-ph.CO].

\bibitem{Gunn:1972sv}
  J.~E.~Gunn and J.~R.~I.~Gott,
  ``On the infall of matter into cluster of galaxies and some effects on their evolution,''
  Astrophys.\ J.\  {\bf 176}, 1 (1972).

\bibitem{Press:1973iz}
  W.~H.~Press and P.~Schechter,
  ``Formation of galaxies and clusters of galaxies by selfsimilar gravitational
  condensation,''
  Astrophys.\ J.\  {\bf 187}, 425 (1974).

\bibitem{Bond:1990iw}
  J.~R.~Bond, S.~Cole, G.~Efstathiou and N.~Kaiser,
  ``Excursion set mass functions for hierarchical Gaussian fluctuations,''
  Astrophys.\ J.\  {\bf 379}, 440 (1991).

\bibitem{Matarrese:2000iz}
  S.~Matarrese, L.~Verde and R.~Jimenez,
  ``The abundance of high-redshift objects as a probe of non-Gaussian initial conditions,''
  Astrophys.\ J.\  {\bf 541} (2000) 10
  [arXiv:astro-ph/0001366].

\bibitem{LoVerde:2007ri}
  M.~LoVerde, A.~Miller, S.~Shandera and L.~Verde,
  ``Effects of Scale-Dependent Non-Gaussianity on Cosmological Structures,''
  JCAP {\bf 0804} (2008) 014
  [arXiv:0711.4126 [astro-ph]].

\bibitem{Sheth:2001dp}
  R.~K.~Sheth and G.~Tormen,
  ``An Excursion Set Model Of Hierarchical Clustering : Ellipsoidal Collapse
  And The Moving Barrier,''
  Mon.\ Not.\ Roy.\ Astron.\ Soc.\  {\bf 329}, 61 (2002)
  [arXiv:astro-ph/0105113].

\bibitem{Maggiore:2009rv}
  M.~Maggiore and A.~Riotto,
  ``The Halo Mass Function from the Excursion Set Method. I. First principle
  derivation for the non-Markovian case of Gaussian fluctuations and generic
  filter,''
  arXiv:0903.1249 [astro-ph.CO].

\bibitem{Maggiore:2009rw}
  M.~Maggiore and A.~Riotto,
  ``The halo mass function from the excursion set method. II. The diffusing
  barrier,''
  arXiv:0903.1250 [astro-ph.CO].

\bibitem{Maggiore:2009rx}
  M.~Maggiore and A.~Riotto,
  ``The halo mass function from the excursion set method. III. First
  principle derivation for non-Gaussian theories,'' 
  arXiv:0903.1251 [astro-ph.CO].

\bibitem{Valageas:2009vn}
  P.~Valageas,
  ``Mass function and bias of dark matter halos for non-Gaussian initial
  conditions,''
  arXiv:0906.1042 [astro-ph.CO].


\bibitem{Bardeen:1985tr}
  J.~M.~Bardeen, J.~R.~Bond, N.~Kaiser and A.~S.~Szalay,
  ``The Statistics Of Peaks Of Gaussian Random Fields,''
  Astrophys.\ J.\  {\bf 304}, 15 (1986).

\bibitem{Sheth:1999su}
  R.~K.~Sheth, H.~J.~Mo and G.~Tormen,
  ``Ellipsoidal collapse and an improved model for the number and spatial
  distribution of dark matter haloes,''
  Mon.\ Not.\ Roy.\ Astron.\ Soc.\  {\bf 323}, 1 (2001)
  [arXiv:astro-ph/9907024].

\bibitem{Lam:2009nb}
  T.~Y.~Lam and R.~K.~Sheth,
  ``Halo abundances in the \fnl model,''
  arXiv:0905.1702 [astro-ph.CO].


\bibitem{Sugiyama:1994ed}
  N.~Sugiyama,
  ``Cosmic background anistropies in CDM cosmology,''
  Astrophys.\ J.\ Suppl.\  {\bf 100}, 281 (1995)
  [arXiv:astro-ph/9412025].

\bibitem{Seljak:1996is}
  U.~Seljak and M.~Zaldarriaga,
  ``A Line of Sight Approach to Cosmic Microwave Background Anisotropies,''
  Astrophys.\ J.\  {\bf 469} (1996) 437
  [arXiv:astro-ph/9603033].

\bibitem{Lewis:1999bs}
  A.~Lewis, A.~Challinor and A.~Lasenby,
  ``Efficient Computation of CMB anisotropies in closed FRW models,''
  Astrophys.\ J.\  {\bf 538} (2000) 473
  [arXiv:astro-ph/9911177].

\bibitem{Babich:2004gb}
  D.~Babich, P.~Creminelli and M.~Zaldarriaga,
  ``The shape of non-Gaussianities,''
  JCAP {\bf 0408} (2004) 009
  [arXiv:astro-ph/0405356].

\bibitem{Lyth:2002my}
  D.~H.~Lyth, C.~Ungarelli and D.~Wands,
  ``The primordial density perturbation in the curvaton scenario,''
  Phys.\ Rev.\  D {\bf 67} (2003) 023503
  [arXiv:astro-ph/0208055].

\bibitem{Bartolo:2003jx}
  N.~Bartolo, S.~Matarrese and A.~Riotto,
  ``On non-Gaussianity in the curvaton scenario,''
  Phys.\ Rev.\  D {\bf 69} (2004) 043503
  [arXiv:hep-ph/0309033].

\bibitem{Dvali:2003em}
  G.~Dvali, A.~Gruzinov and M.~Zaldarriaga,
  ``A new mechanism for generating density perturbations from inflation,''
  Phys.\ Rev.\  D {\bf 69} (2004) 023505
  [arXiv:astro-ph/0303591].
  
\bibitem{Alishahiha:2004eh}
  M.~Alishahiha, E.~Silverstein and D.~Tong,
  ``DBI in the sky,''
  Phys.\ Rev.\  D {\bf 70}, 123505 (2004)
  [arXiv:hep-th/0404084].
  
\bibitem{ArkaniHamed:2003uz}
  N.~Arkani-Hamed, P.~Creminelli, S.~Mukohyama and M.~Zaldarriaga,
  ``Ghost Inflation,''
  JCAP {\bf 0404}, 001 (2004)
  [arXiv:hep-th/0312100].
    
\bibitem{Creminelli:2003iq}
  P.~Creminelli,
  ``On non-Gaussianities in single-field inflation,''
  JCAP {\bf 0310}, 003 (2003)
  [arXiv:astro-ph/0306122].

\bibitem{Creminelli:2005hu}
  P.~Creminelli, A.~Nicolis, L.~Senatore, M.~Tegmark and M.~Zaldarriaga,
  ``Limits on non-Gaussianities from WMAP data,''
  JCAP {\bf 0605} (2006) 004
  [arXiv:astro-ph/0509029].

\bibitem{ZinnJustin:2002ru}
  J.~Zinn-Justin,
  ``Quantum field theory and critical phenomena,''
  Int.\ Ser.\ Monogr.\ Phys.\  {\bf 113} (2002) 1.

\bibitem{Chandrasekhar:1943ws}
  S.~Chandrasekhar,
  ``Stochastic problems in physics and astronomy,''
  Rev.\ Mod.\ Phys.\  {\bf 15}, 1 (1943).

\bibitem{Giannantonio:2009ak}
  T.~Giannantonio and C.~Porciani,
  ``Structure formation from non-Gaussian initial conditions:
  multivariate biasing, statistics, and comparison with N-body
  simulations,'' 
  arXiv:0911.0017 [astro-ph.CO].

\bibitem{Carbone:2009zz}
  C.~Carbone {\it et al.},
  ``The properties of the dark matter halo distribution in non-Gaussian
  scenarios,''
  Nucl.\ Phys.\ Proc.\ Suppl.\  {\bf 194}, 22 (2009).

\bibitem{Bernardeau:2001qr}
  F.~Bernardeau, S.~Colombi, E.~Gaztanaga and R.~Scoccimarro,
  ``Large-scale structure of the universe and cosmological perturbation
  theory,''
  Phys.\ Rept.\  {\bf 367}, 1 (2002)
  [arXiv:astro-ph/0112551].

\bibitem{Robertson:2008jr}
  B.~Robertson, A.~Kravtsov, J.~Tinker and A.~Zentner,
  ``Collapse Barriers and Halo Abundance: Testing the Excursion Set Ansatz,''
  Astrophys.\ J.\  {\bf 696} (2009) 636
  [arXiv:0812.3148 [astro-ph]].



\bibitem{Sheth:2003py}
  R.~K.~Sheth and R.~van de Weygaert,
  ``A hierarchy of voids: Much ado about nothing,''
  Mon.\ Not.\ Roy.\ Astron.\ Soc.\  {\bf 350}, 517 (2004)
  [arXiv:astro-ph/0311260].

\bibitem{Kamionkowski:2008sr}
  M.~Kamionkowski, L.~Verde and R.~Jimenez,
  ``The Void Abundance with Non-Gaussian Primordial Perturbations,''
  JCAP {\bf 0901}, 010 (2009)
  [arXiv:0809.0506 [astro-ph]].

\bibitem{Erdelyi}
  A.~Erd\'elyi,
  ``Asymptotic Expansions,''
  {\it New York, NY: Dover (1956) 108pp.}
\end{thebibliography}
\end{document}